\documentclass[twocolumn,superscriptaddress,secnumarabic,amssymb,nobibnotes,aps,pra]{revtex4-1}

\usepackage{extarrows}
\usepackage{amsmath}    
\usepackage{graphicx}   
\usepackage{verbatim}   
\usepackage{color}      
\usepackage{subfigure}  
\usepackage{hyperref}   
\usepackage{verbatim}   
\usepackage{longtable}
\usepackage{epsfig}
\usepackage{comment}
\usepackage{bm}
\usepackage{ulem}
\usepackage{dcolumn} 
\usepackage{textcomp}   
\usepackage{kotex}      
\newcommand\redout{\bgroup\markoverwith
	{\textcolor{red}{\rule[0.5ex]{2pt}{0.8pt}}}\ULon}

\newcommand{\noo}{NaOsO$_3$}
\newcommand{\tr}{$t_\frac{1}{2}$}

\begin{document}
\title{Altermagnetism and Weak Magnetism \\
in the Insulating Distorted Perovskite Antiferromagnet NaOsO$_3$}
\author{Hong-Suk Choi}
\affiliation{Department of Applied Physics, Graduate School, Korea University, Sejong 30019, Korea}
\author{Myung-Chul Jung}
\affiliation{Division of Semiconductor Physics, Korea University, Sejong 30019, Korea}
\author{Kyo-Hoon Ahn}
\email{kyohoon.ahn@fzu.cz}
\affiliation{Institute of Physics, Czech Academy of Sciences, Cukrovarnick\'a 10, 162 00 Praha 6, Czechia}
\author{Warren E. Pickett}
\email{wepickett@ucdavis.edu}
\affiliation{Department of Physics and Astronomy, University of California Davis, California 95616, USA }
\author{Kwan-Woo Lee}
\email{mckwan@korea.ac.kr}
\affiliation{Department of Applied Physics, Graduate School, Korea University, Sejong 30019, Korea}
\affiliation{Division of Semiconductor Physics, Korea University, Sejong 30019, Korea}
\date{\today}

\begin{abstract}
The GdFeO$_3$-type perovskite antiferromagnet NaOsO$_3$, calculated here to be altermagnetic for all three typical collinear antiferromagnetic orders,
was suggested early on to be a Slater-type insulator, due in large part to its continuous metal-insulator transition and its small energy gap.
Below the N\'eel temperature, the gap opens along with ``weak magnetism'', accompanied by a 
sharp change in the magnetic susceptibility and resistivity.
Without explicit correlation in the band structure calculation, and neglecting spin-orbit coupling (SOC), 
already a gap opens.
Inclusion of a modest on-site Coulomb repulsion ($U\sim$1 eV) is sufficient to eliminate a 
SOC-induced small band overlap, 
opening a gap similar to the experimentally observed gap of around 100 meV. 
Combined evidence supports the viewpoint that \noo~ lies in an unusual crossover region between Slater and Mott insulator. 
The unreported altermagnetism in \noo~is demonstrated and its consequences are considered.
The origin of the very weak magnetism 
has been investigated using a combination of {\it ab initio} calculations and symmetry analysis of 
the magnetic space group, confirming the origin lying in the Dzyaloshinskii-Moriya SOC 
buttressed by altermagnetic order. After determining the easy axis, our calculation 
leads to an Os spin canting angle of about 3$^{\circ}$, 
accounting for the observed weak magnetism and sharp change in the susceptibility.  
The altermagnetism spin-split bands (up  to $\sim$100 meV) are accompanied by a chiral-split 
magnon spectrum in both acoustic and optical modes in the THz range, 
and lead to significant anomalous Hall conductivity upon hole doping.
\end{abstract}

\maketitle

\clearpage
\section{Introduction}
Recently altermagnetism has become an active topic in condensed matter physics,
since altermagnets (AMs) display unexpected properties, viz. spin-split
bands in a zero net moment antiferromagnet (AFM), 
with or without the help of spin-orbit coupling (SOC). 
Other properties,  observed in various topological systems
as well as ferromagneticlike features with no stray field, 
show promising potential for spintronic applications \cite{prx1,prx2,rev1,rev2,j.Liu2021}.
In conventional AFMs, spin-opposite sublattices 
are mapped by translation $t$ or inversion $P$ symmetries, 
leading to Kramers degeneracy throughout the Brillouin zone (BZ). 
However, in AMs the AFM sublattice sites are connected by a combination of 
rotational $C_{nj}$ ($n$-fold rotation about the $\hat{j}$-axis) 
and nonprimitive translation symmetries. As a result, the Kramers degeneracy is broken 
in some regions of BZ \cite{c.Wu2007,ruo2KU,hayami2019,yuan2020,libor2020,mazin2021,cheong2024,krem2024}, 
leading to spin-split bands in the selected regions in spite of vanishing net moment.
After this phenomenon was theoretically reported in rutile AFM RuO$_2$ 
by Ahn, Hariki, Lee, and Kune\v s in 2019 \cite{ruo2KU}, 
various unique features of this novel AFM have been systematically investigated by 
\v Smejkal, Sinova, and Jungwirth in 2022 \cite{prx1,prx2},
after which the phenomenon became known as altermagnetism.

Another particularity is that spin-splittings in the AM lead to chiral magnons 
that are anticipated to be a useful new property 
for applications to magnonics due to linear dispersion and the high energy 
scale of THz \cite{chiralmagnon,liu2024}.
Spontaneous crystal thermal Hall effects are proposed in both metallic or 
insulating AMs \cite{zhou2024,hoyer2025},
suggesting promising candidates for application to spin caloritronics.
Additionally, the crystal chirality in a conducting AM induces anomalous Hall 
conductivity (AHC) \cite{ahe_rev2022,feng2022,ahe2023,ahe2024}, 
a property that is more often associated with metallic ferromagnets.

In this paper we focus on the GdFeO$_3$-type, 
quadrupled orthorhombic perovskites, \noo~in particular.
Perovskites typically exhibit three possible collinear 
AFM spin orders: G type (antialigned neighboring spins), 
C type (aligned along one direction and antialigned along a perpendicular 
direction), and A type (alternatively antialigned layers).
In these orderings, the magnetic sublattices are connected by a combination 
of two-fold rotational $C_{2j}$
and non-primitive translation \tr~ symmetries characteristic of AMs [see Supplemental Material (SM) \cite{sm}, for details].
The crystal structure and AFM alignment dictate that AFM GdFeO$_3$-type perovskites are AMs
in all of these collinear AFM orders \cite{perovskiteAM}. 
In this paper we provide 
direct evidence that AFM \noo~possesses spin-split bands and chiral magnons,
the two most evident properties of AMs, and related properties are discussed.

Over last 15 years, GdFeO$_3$-type \noo~ has been investigated  
on both experimental \cite{shi2009,calder2012,lupi2013,calder2017,gurung2018,vale2018prl,vale2018prb,sereika2020} 
and theoretical sides \cite{savrasov,nooKU,middey2014,bkim2016,moha2018,ntallis2021,antonov2022,johannes2023,bhandari2019},
with an overview given in the next section. 
Perhaps the most unusual feature on NaOsO$_3$ is that it was suggested \cite{shi2009} to be a promising 
candidate for a continuous Slater metal-insulator transition (MIT) \cite{slater}
where a gap opens due to antiferromagnetic order with no relevant change of
the crystal, either symmetry or internal structure. Previous theoretical 
work (see the next section) has shown that
correlation corrections, though not large, are necessary to account for 
the insulating and AFM ground state that is observed. Correlations argue for a
Mott transition involving correlation, Slater's original picture \cite{slater} was based 
on a Hartree-Fock treatment with exchange being the underlying interaction. 
Since the appropriate picture of this transition is of
primary interest in NaOsO$_3$, it is worthwhile to consider this rather high 
temperature MIT somewhat more broadly.

Some time ago Halperin and Rice provided a survey of such transitions 
from the point of view of a MIT with increasing 
temperature \cite{halperin1,halperin2}, arising
from a (most interestingly, indirect) bandgap closing, which seems to
fit the current theoretical picture and much of the experimental
data. At that time, their picture of
a Mott transition was of a localized-to-itinerant electronic system,
which would be first order (``no phase transition"). The Slater transition
would be continuous and second order, the AFM order disappearing
continuously to zero at the transition, unlike the discontinuous
AFM order parameter at a Mott transition. Observed MITs have become
much more complicated than these simple pictures as more interactions
are included, viz. the temperature dependence of band energies due
to electron-phonon coupling \cite{Allen-Cardona1983}, and more complex compounds
and active multiorbital electronic structures have produced ever
richer behaviors \cite{Imada1998}.

However, the main emphasis of Halperin and Rice was other effects,
of interactions and temperature. Assuming sharp band structures and
a well defined gap E$_g$(T), they argued that (especially for indirect
gaps) the electron-hole attraction would lead to an excitonic phase
before the gap vanished. A great deal of research has been conducted
on excitonic behavior in the meantime. There are other important
effects they did not consider. The electron-phonon interaction, often
strong in transition-metal oxides and not considered by them, leads to renormalization 
of the bands and gap from the band value,\cite{Allen-Heine1976,Allen-Cardona1983} as well
as providing lifetime effects thus broadening bands and somewhat 
smoothing the gap, and making the description of a sharp MIT (whether
first or second order) a more involved task.

In this paper, after demonstrating band splitting and chiral magnons,
we will focus on these questions of importance:
(1) how does AM figure into these phenomena or the properties of \noo?
(2) what is the origin of the observed very weak magnetization?
(3) how might one best consider the Slater versus Mott insulator question?
The organization of the paper is as follows. Section II provides an overview
of experimental data and theoretical modeling of NaOsO$_3$, with our Methods 
described in Sec. III. The main results 
from band modeling, on magnonic properties, the source of the weak magnetism, 
and anomalous Hall effect in the doped system are provided in several
subsections of Sec. IV. The 
paper concludes with a Discussion and Summary in Sec. V.

\section{Survey of data and previous theory}
Shi {\it et al.} \cite{shi2009} discovered that \noo~ becomes AFM at the relatively high 
N\'eel temperature of $T_N=410$ K and 
correspondingly large Curie-Weiss temperature of $\Theta_{CW}=-1950$ K.
This was coincident with a continuous opening of a gap \cite{shi2009}.
Neutron powder diffraction (NPD) measurements by Calder {\it et al.}
indicated G-type AFM with 
the easy axis along the $\hat{c}$-direction \cite{calder2012},
confirmed by resonant inelastic x-ray scattering (RIXS) \cite{calder2017}
and several calculations based on density functional theory 
(DFT) \cite{savrasov,nooKU,middey2014,antonov2022,ntallis2021}.
Vecchio {\it et al.} used infrared and THz spectroscopy to monitor the
energy gap \cite{lupi2013}, 
giving a value of the low temperature optical gap of $\sim$ 105 meV 
and a fully gapped Fermi surface .
At the MIT the crystal symmetry and the volume remained 
unchanged  \cite{shi2009,calder2012,calder2017,gurung2018,vale2018prl,vale2018prb},
supporting the suggestion of Shi {\it et al.} \cite{shi2009} that the AFM of \noo~ is due to Slater transition.

Through RIXS, the spin gap was measured as 58--60 meV \cite{calder2017,vale2018prl,vale2018prb}.
Measurements of non-resonant and resonant x-ray single-crystal diffraction at the Os L$_3$ edges
showed the space-group forbidden (300) Bragg peak as well as the magnetic (330) 
peak emerged at $T_N$ \cite{gurung2018}.
The forbidden peak was interpreted as the change in the electron density of Os,
suggested to support the Slater mechanism \cite{gurung2018}.
High pressure experiments indicate that this small gap insulating state remains robust up to 35 GPa 
with the MIT critical temperature dropping to $T_c\approx300$ K\cite{sereika2020}. 

Vale {\it et al.} observed the magnetic excitation through RIXS that was weakly 
Landau-damped upon approaching the metallic region \cite{vale2018prl,vale2018prb}.
This feature was well described by a weak AFM Fermi liquid picture, suggesting
that \noo~is not well suited to the standard Slater picture \cite{vale2018prl,vale2018prb}.
On the other hand, the Bragg reflections 
were well reproduced with uncorrelated {\it ab initio} calculations, 
suggesting that this aspect of \noo~can be accounted for within the Slater picture \cite{gurung2018}.

Several theoretical works based on the DFT have appeared, although nothing about the AM aspect.
Several DFT+SOC+U calculations established SOC to be
important, and proposing that an effective Coulomb repulsion 
$U_{eff}=U-J_H$ somewhat larger than 1 eV (mostly $U_{eff}\approx2$ eV)
seemed appropriate to account for the MIT \cite{savrasov,nooKU,antonov2022,johannes2023},
implying that correlation, but only a modest amount, is necessary 
to account for the MIT. Here $J_H$ is the conventional Hund's exchange integral.
Zhou {\it et al.} pointed out
the small change in the slope of the temperature dependence 
of the lattice parameter difference $|c-a|$, with onset below
$T_{N}$ $(=T_{MIT}) $\cite{johannes2023},
although no change in volume versus temperature was observed. 
They suggested that this system is close to the Mott physics picture 
in spite of the very small gap.
Alternatively, Kim {\it et al.} suggested 
a magnetically induced continuous change in the topology of the Fermi surface, called a spin-driven Lifshitz MIT, based on the coexistence of weak correlation and itinerant magnetism \cite{bkim2016}.

Measurements of isothermal magnetization showed 
a weak spontaneous magnetization of 0.005$\mu_B$/Os at 5 K  \cite{shi2009},
while no evidence for spin canting was detected in the NPD Bragg peaks \cite{calder2012}.
Although magnetic frustration is not an issue in perovskites, 
the unusually high Curie-Weiss temperature can be used to extract the 
empirical frustration factor, defined by Ramirez \cite{ramirez}, 
of $\frac{\Theta_{CW}}{T_N}\approx$4.8.
This value is comparable with that of the magnetically frustrated system 
YMnO$_3$ \cite{ramirez}, 
leaving this large ratio as an open question. 
Calculations based on DFT of non-collinear magnetic configurations 
indicated that G-type non-collinear ordering is energetically favored by
several meV over collinear G-type ordering \cite{antonov2022}. 
Currently the origin of the observed weak magnetism remains elusive.

\section{Methods}
We have used the experimentally observed lattice parameters of $a=5.3842$, 
$b=7.5804$, $c=5.3282$ (in units of \AA) 
in the GdFeO$_3$-type $\sqrt{2}\times2\times\sqrt{2}$ perovskite structure (space group: No. 62 $Pnma$, with
generators of inversion and two non-symmorphic operations) \cite{shi2009},
as used in the previous paper by some of the current authors \cite{nooKU}.
In this distorted perovskite structure, an O--Os--O axis rotates by 
9$^\circ$ about the $\hat{a}$ ($\hat{c}$) axis and tilts by 11$^\circ$ 
away from the $\hat{b}$ axis. 
Although the distortions are moderate, they play a crucial role in the 
emergence of the AFM/MIT transition, as shown in previous calculations \cite{nooKU,moha2018,bhandari2019}, 
and hence the low temperature AM state.

Our {\it ab initio} calculations, based on the generalized gradient 
approximation (GGA) of Perdew, Burke, and Ernzerhof \cite{gga}, 
were carried out with the Vienna ab initio simulation package ({\sc vasp}) \cite{vasp1,vasp2}. 
An energy cutoff of 600 eV was employed for the plane-wave basis set, and the BZ was sampled 
by a $16\times12\times16$ Monkhorst-Pack $k$-point mesh \cite{mp}.
SOC and correlation effects were treated within 
GGA+SOC, GGA+U, and GGA+U+SOC approaches, as
implemented in {\sc vasp}. Correlation effects in the half-filled $t_{2g}$ 
shell were included by the effective on-site Coulomb repulsion $U_{eff}=U-J_H$,
using the scheme proposed by Dudarev {\it et al.} \cite{sutton} 
that depends only on $U_{eff}$.
We varied $U_{eff}$ in the 0.5--2 eV range to include correlation effects up to the value that is reasonable for Os ions in the small-gap semiconductor.
As expected, increasing $U_{eff}$ enlarges the energy gap 
and slightly varies the magnitude of the parameters studied here. 
However, the characteristic features induced by AM, 
{\it e.g.} chiral-split magnon and significant AHC around the Fermi energy $E_F$,
remain almost unchanged.
In particular, compared to the existing experimental data,
the choice of $U_{eff}=0.5$ eV in the GGA+U+SOC calculations  
well reproduces the energy gap and the (maximum) values of the magnon spectrum,
as will be shown below.
Thus, we have focused mainly on results for $U_{eff}=0.5$ eV.

Maximally localized Wannier functions \cite{marzari97} were constructed 
using the {\sc wannier90} program \cite{wan90}. 
For the Wannierization, we used 56 orbitals including all of the Os $5d$ 
and O $2p$ orbitals in the quadrupled perovskite cell,
showing a good fit in the range of -9 eV to +7eV \cite{sm}.
The symmetries of the Wannier Hamiltonian obtained above were analyzed, 
using the {\sc wannsymm} code \cite{wansym}.
From the symmetrized Hamiltonian, the magnetic exchange parameters 
and the Dzyaloshinskii-Moriya interaction (DMI) constants \cite{dm1,dm2} 
were calculated using the {\sc tb2j} package \cite{tb2j}
in which the local rigid spin rotation is treated as a perturbation 
within the single-particle Green's function method \cite{liecht}.

Using the exchange constants, the spin dynamics was simulated by 
the {\sc uppasd} program \cite{uppasd}.
For rigid band doping, the AHC can be calculated from {\it ab initio} 
calculations of the Berry curvature using the Wannier Hamiltonian, implemented in the {\sc wanniertools} code \cite{wantool}.

\section{Results}
\subsection{Energetics and gap opening}

\begin{figure}[htp] 
  \includegraphics[width=\columnwidth]{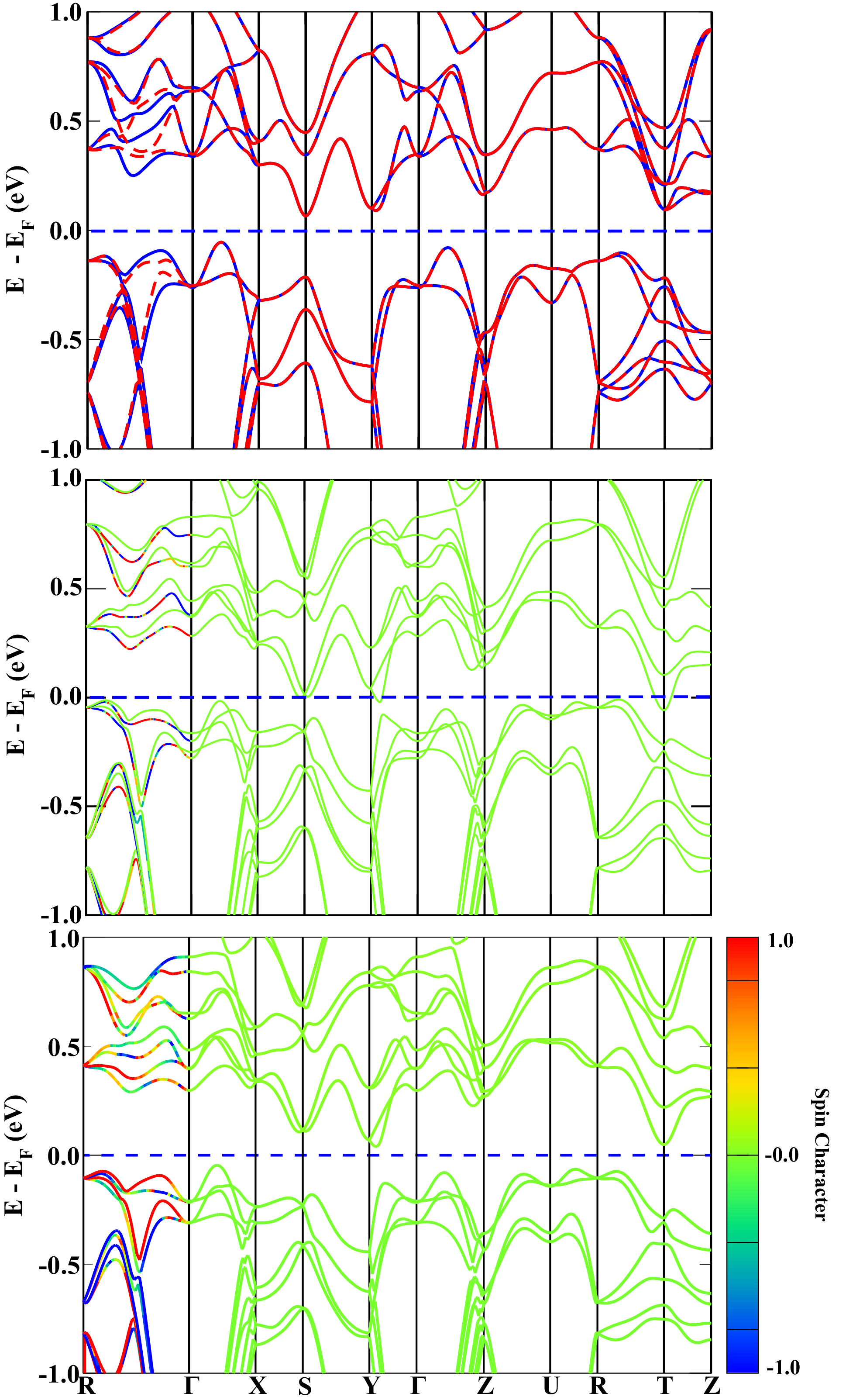}
\caption{Enlarged band structures of \noo~ with (top) GGA, (middle) GGA+SOC, and (bottom) GGA+SOC+$U$ 
near $E_F$.
Here, $U_{eff}$=0.5 eV and the N\'eel vector ${\bf N}\|\hat{z}$ were used.
The spin characters are colored (red: spin-up, blue: spin-down), revealing the
spin splittings along the $R-\Gamma$ line, indicating the altermagnetism character.
Note that the splitting occurs only along the $R-\Gamma$ line (and inside the zone), but not along other symmetry directions.
(For details, see Sec. IV \ref{spin-splitting}.)
Band sticking is discussed in the text.
}
\label{bs}
\end{figure}

To verify our results obtained from calculations based on the GGA, 
first we investigated energetics and compared with the experimental 
measurements and previous theoretical results. 
Consistent with the literature \cite{calder2012,calder2017,savrasov,nooKU,middey2014,antonov2022}, 
the AFM state has a much lower energy than the nonmagnetic and ferromagnetic states.
At the GGA level, the G-AFM is energetically favored over the A-type by 36 meV/f.u. and the C-type by 55 meV/f.u.,
consistent with the previous calculations \cite{nooKU}.
To find the easy axis, we carried out GGA+SOC calculations for three directions of 
the N\'eel vector {\bf  N}=${\bf m}_1-{\bf m}_2$ 
in terms of the two sublattice magnetic vectors \cite{ahe_rev2022}, until SOC is added. 
Consistent with the experimental observations \cite{calder2012,calder2017},
the ${\bf N}\|\hat{z}$ state has a little lower energy than in the ${\bf N}\|\hat{x}$  and ${\bf N}\|\hat{y}$ states
by 3.2 meV and 4.1 meV per f.u., respectively.
Including the correlation effects, the energetics remains unchanged.

Figure \ref{bs} shows the band structures enlarged around $E_F$ 
for the GGA, GGA+SOC(${\bf N}\|\hat{z}$), and GGA+SOC(${\bf N}\|\hat{z}$)+U with $U_{eff}=0.5$ eV.
In contrast to our previous results of the local spin density approximation (LSDA)  \cite{nooKU}, 
pure GGA opens a gap $E_g$ of 150 meV for the AFM state without a help of $U$.
It is rare for these two approaches, the GGA and the LSDA, to show opposite results 
regarding whether a system is metallic or insulating.
The reason is unclear, but highlights yet another peculiarity of \noo.
Moreover, including SOC a small negative indirect gap around 50 meV appears between the $R$ and $T$ points
in the middle panel of Fig. \ref{bs}, although the valence and conduction bands remain disjoint.
After this, inclusion of a small on-site Coulomb repulsion is sufficient to re-open a gap.
At $U_{eff}=0.5$ eV, 
our calculated gaps are 29, 2, 70 (in units of meV) for ${\bf N}\|\hat{x}$, 
$\|\hat{y}$, and $\|\hat{z}$, respectively.
For the easy axis ${\bf N}\|\hat{z}$, the energy gap is consistent with the 
experimental value of 60--100 meV.
At $U_{eff}=1 (2)$ eV, the energy gap is about 0.2 (0.5) eV,
at least twice larger than the experimental value.
Thus, now we will focus on the results obtained from 
GGA+SOC(${\bf N}\|\hat{z}$)+U at $U_{eff}=0.5$ eV, unless mentioned otherwise.
Considering a Hund's integral $J_H\approx 0.5$ eV typical for osmates \cite{nooKU},
this $U_{eff}$ corresponds to the Hubbard $U\approx1$ eV,
around half of that indicated by previous DFT 
calculations \cite{savrasov,nooKU,antonov2022,johannes2023}.

In the band structure plots enlarged in the --1 eV to 1 eV in Fig. \ref{bs}, 
there is band sticking along certain symmetry lines.
Bands ``stick together'' at the top of the BZ, along the $Z-U-R-T-Z$ line, 
a common feature of non-symmorphic crystals \cite{jeong}. 
However, bands also stick together along the $X-S-Y$ line in the basal plane, due to the second
non-symmorphic group generator.

\begin{figure}[tbp] 
\includegraphics[width=\columnwidth]{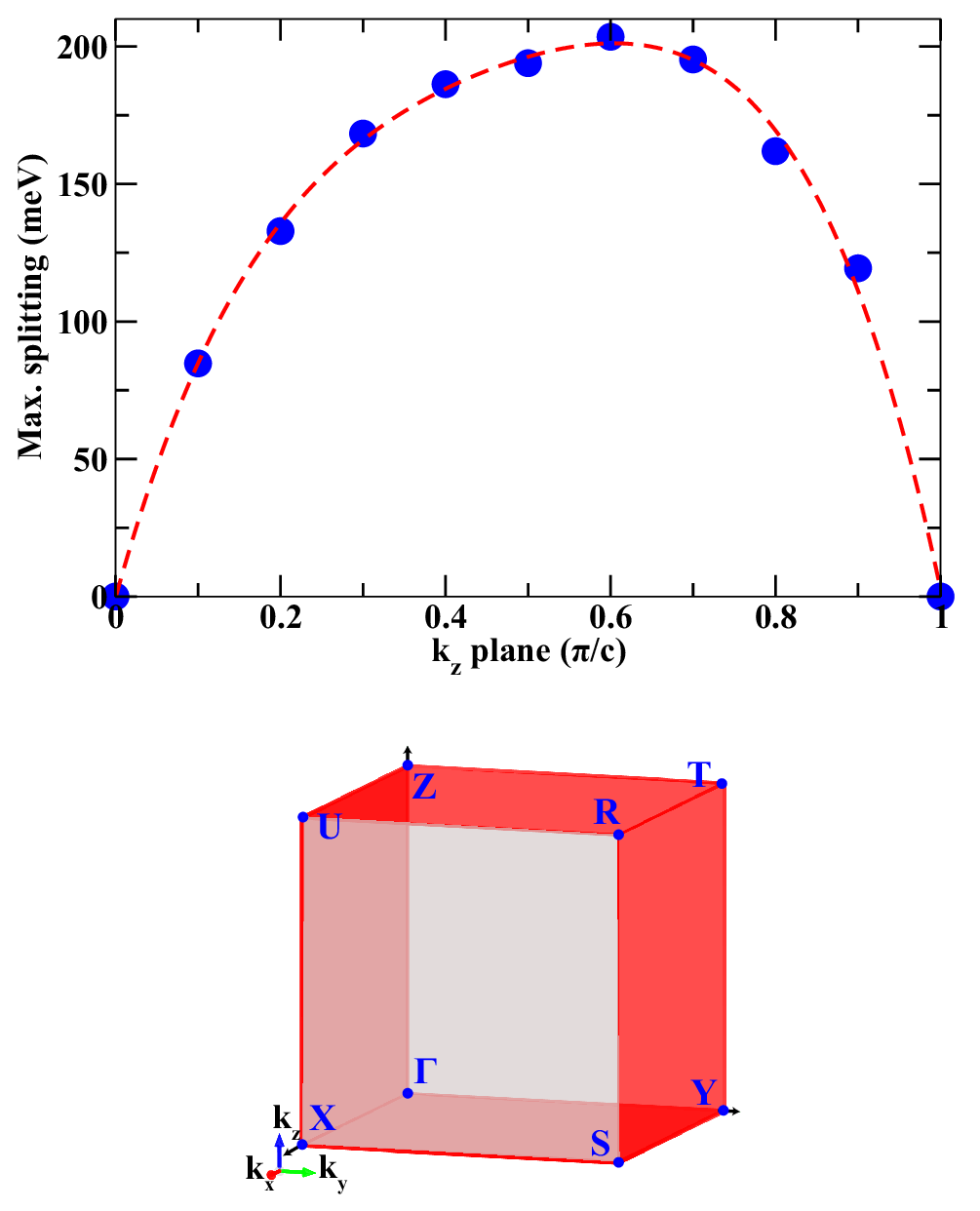} 
\caption{Top: maximum splitting on the $k_x-k_y$ planes, 
for $k_z$ varying through the range of $0-\frac{\pi}{c}$.
The variation in splittings is approximately quartic,  
as indicated from a fit to the data and shown by the red 
dashed lines (see text for more discussion).
Bottom: sketch of the region showing spin-splitting in the orthorhombic BZ.
In the red-shaded or lined regions, the Kramers spin-degenerate bands survive. 
}
\label{splitting}
\end{figure}

\subsection{Spin-splitting induced by altermagnetism}
\label{spin-splitting}
An early example of AM-induced band splitting in an AFM was demonstrated
in RuO$_2$ by Ahn {\it et al.} \cite{ruo2KU}
In the \noo~band structures displayed in Fig. \ref{bs}, 
this characteristic feature of AM appears, recall that conventional
AFMs have degeneracy related by reversal of spins.
Clearly visible are spin-splittings along the $R-\Gamma$ line, and
at the top of the valence bands and the bottom of the conduction bands,
within the range of around $-1$ eV to 1 eV.
The characters of bands in this region are mostly the Os $5d$ orbitals,
and band-resolved magnetization is evident.
Since in  other energy ranges the splittings are negligible,
this implies that the magnitude of splitting significantly depends 
on strength of the exchange interaction in specific bands.

As far as we know, no established way exists 
for determining the magnitude of spin splitting, yet.
We scanned the entire BZ to investigate the magnitude of splitting 
at a certain $\vec{k}$-point,
since its magnitude varies across the BZ.
In \noo~ which has the spin space group of $P^1n^{-1}m^{-1}a^{\infty m}1$ \cite{ssg1,ssg2},
the Kramers spin degeneracy survives on the $k_z(k_y)=0$ and 
$\frac{\pi}{c}(\frac{\pi}{b})$ planes
due to the symmetry of $\{2_\|\|m_{001}|t_{\frac{1}{2}0\frac{1}{2}}\}$
($\{m_\perp\|m_{010}|t_{0\frac{1}{2}0}\}$), {\it i.e.}, spin reversing operation and a mirror reflection about the $z$ ($y$)-axis.
Thus, the spin-splittings appear only inside the BZ and on the $k_x=0,\frac{\pi}{a}$ planes \cite{sm},
as sketched in the bottom panel of Fig. \ref{splitting}.

Using bands at the GGA level, Fig.~\ref{splitting} presents the maximum splitting for each $k_x-k_y$ plane, 
varying $k_z$ between 0 and $\frac{\pi}{c}$. 
On most planes, the maximum splitting appears accidentally 
in the midway of the $[01k_z]-[00k_z]$ line. (For details, see the SM \cite{sm}.)
At the GGA level the largest splitting is about 200 meV, 
occurring at the top of valence bands at the $k_z\approx0.6\frac{\pi}{c}$.
With GGA+SOC+U these reduce to about a half of the GGA value.
Compared with the splittings in prototype AMs RuO$_2$, MnTe and CrSb, 
that of GGA in \noo~ is about one fifth of those, but is larger than or comparable to 
other candidate AMs \cite{prx1,prx2}. 
The plotted maximum 
vanishes at the endpoints due to symmetry, and is approximately quadratic
with some displacement of the apex from the center.

\subsection{Weak magnetism}
\label{WM}

\begin{table}[b]
\caption{Magnetic space groups (MSGs) and symmetries for each direction of the N\'eel vector ${\bf N}$ in \noo.
All directions have in common the identity $I$ and inversion $P$ symmetries (not given here).
No additional time reversal symmetry operation $T$ appears in ${\bf N}\|\hat{x}$, 
whereas this symmetry is involved for the other two directions.
}
\begin{center}
\begin{tabular}{cccl}\hline\hline
Dir.\bf{N}  & MSG &~& \multicolumn{1}{c}{symmetries}\\ \hline
$\hat{x}$ &  $P_{nma}.1$ &~& $C_{2z}$\tr, $C_{2z}$\tr$P$; $C_{2x}$\tr, $C_{2x}$\tr$P$; \\
                ~  & (No. 62.441) &~ & $C_{2y}$\tr, $C_{2y}$\tr$P$ \\
$\hat{y}$ & $P_{n'm'a}$ &~&$C_{2z}$\tr,  $C_{2z}$\tr$P$; $C_{2x}$\tr$T$,  $C_{2x}$\tr$PT$;\\
            ~       & (No. 62.446)   &~& $C_{2y}$\tr$T$,  $C_{2y}$\tr$PT$ \\
$\hat{z}$ & $P_{n'ma'}$ &~&$C_{2z}$\tr$T$, $C_{2z}$\tr$PT$;  $C_{2x}$\tr$T$, $C_{2x}$\tr$PT$;\\
             ~       & (No. 62.448)   &~& $C_{2y}$\tr, $C_{2y}$\tr$P$ \\\hline\hline 
\end{tabular}
\end{center}
\label{table1}
\end{table}

SOC induces spin canting, and hence weak magnetism (WM) in distorted AFMs 
via the Dzyaloshinskii-Moriya mechanism. 
Presence of this WM can be analyzed through symmetries of the magnetic space groups 
that depend on the direction of the N\'eel vector.
Neglecting SOC, 
each magnetic space group has symmetries of a combination of  rotation $C_{2j}$, 
translation \tr, inversion $P$, and time reversal $T$ operations, as given in Table \ref{table1},
obtained using the {\sc wannsymm} code \cite{wansym}.
In the each space group, the magnetic moment ${\bf  m}$  or Hall vectors ${\bf  h}$ exists
when each vector is invariant for the all symmetry operations of the space group.
These vectors are affected by the inversion $P$ ($\vec{r} \rightarrow -\vec{r}$) 
and the rotation $C_{2j}$  ($\vec{r}_{\perp} \rightarrow -\vec{r}_\perp$, 
components perpendicular to the $\hat{j}$-axis).
On the other hand, the time reversal or translation operator does not change these vectors 
that are quantities obtained from an integration over the BZ.
For a moment ${\bf m}$, for example, operations of $C_{2x}{\bf m}$ and $C_{2x}P{\bf m}$ produce
$(m_x,-m_y,-m_z)$ and $(-m_x,m_y,m_z)$, respectively. 
Since these should be equal to ${\bf m}$ to be invariant, 
$m_y=m_z=0$ and $m_x=0$. Thus non-zero ${\bf m}$ is not allowed in this case.
Therefore, ${\bf m}$ is allowed only along the $\hat{z}$ and $\hat{y}$
for the cases of ${\bf N}\|\hat{y}$ and $\|\hat{z}$,
respectively, whereas ${\bf m}$=0 for ${\bf N}\|\hat{x}$.

\begin{table}[tb]
\caption{Each component (in units of $\mu_B$) of spin and orbital moment vectors 
for GGA+SOC(${\bf N}||\hat{z}$)+$U$, obtained from the {\sc vasp}.
There are two-pairs of antialigned magnetic Os ions in the unit cell.
}
\begin{center}
\begin{tabular}{cccccccc}\hline\hline
~  & \multicolumn{3}{c}{spin} &~& \multicolumn{3}{c}{orbital}\\ \cline{2-4}\cline{6-8}
 ion& $x$ & $y$ & $z$ &~& $x$ & $y$ & $z$ \\\hline
 Os1 & $-0.058$ & $-0.018$ & ~~1.139 &~&~~0.004 & $-0.016$ & $-0.068$ \\
 Os2  & $-0.058$ & $-0.018$ & $-1.139$ &~&~~0.004 & $-0.016$ & ~~0.068 \\
 Os3  & ~~0.058 & $-0.018$ & ~~1.139 &~& $-0.004$ & $-0.016$& $-0.068$ \\
 Os4 & ~~0.058 & $-0.018$ & $-1.139$ &~& $-0.004$ & $-0.016$ & ~~0.068 \\\hline
sum(/ion) & 0  &$-0.018$ &0   &~& 0 & $-0.016$ & 0 \\\hline\hline 
\end{tabular}
\end{center}
\label{table2}
\end{table}

To investigate this WM, GGA+SOC+U calculations were carried out. 
Our results are given in Table \ref{table2}.
The spin moments of the Os ions are 
0.019$\hat{z}$ $\mu_B$ for ${\bf N}\|\hat{y}$ and $-0.018\hat{y}$ $\mu_B$ for ${\bf N}\|\hat{z}$, 
consistent with the symmetry analysis of the magnetic space group.
Considering the total orbital moment of $-0.016$ $\mu_B$,
the total moment is $-0.034\hat{y}$ $\mu_B$/f.u. for ${\bf N}\|\hat{z}$.
This magnitude is very similar to the  observed value of 0.005 $\mu_B$ \cite{shi2009}.
In fact, such a small value is somewhat sensitive to calculational methods (see below).

To follow this in more detail, from Wannierization we obtained a symmetrized Wannier Hamiltonian.
From this Hamiltonian, the spin moments of Os ions were calculated using the {\sc tb2j} code \cite{tb2j}.
For the easy axis ${\bf N}\|\hat{z}$, the calculated local spin moments of the two Os ions 
are ${\bf m}^S_{1,2}=(\pm 0.0521, -0.0182, \mp0.9882) \mu_B$,
leading to ${\bf m}^S={\bf m}_1^S+{\bf m}_2^S=-0.018\hat{y}$ $\mu_B$ per Os ion,
consistent with the value obtained above.
The local spin moment vectors ${\bf m}^S_{1,2}$
indicate a spin canting angle 
$\phi={\rm cos}^{-1}\frac{|m_{1z}^S|}{\|{\bf m}_1^S\|}\approx3.2^\circ$
from the $\hat{c}$-axis, 
not far from the rotation angle of the O--Os--O axis about the axis,
similar to the value of 5.6$^\circ$ obtained from the non-collinear calculations \cite{antonov2022}. 
In this system, the DMI is allowed 
due to broken inversion symmetry by the distortions of oxygen ions 
involved in the magnetic interaction.
This is similar to the AM La$_2$CuO$_4$ \cite{cheong1989}, though in both cases
the magnetic space groups possess inversion symmetry.

Recently, Jo and colleagues suggested 
an intrinsic origin of WM in AM \cite{hwlee2025}.
The anisotropic local symmetry of AM gives rise to anisotropic orbital moments,
which in turn lead to an anisotropic $g$-tensor.
That is, the total moment $\|{\bf m}\|\varpropto\Delta g\approx\frac{L_{\parallel}-L_\perp}{S_0}$
for the two principle axes, where $S_0$ is the magnitude of spin.
Thus,WM can emerge in AM, even in the cases where the DMI is forbidden,
as applied to the case of the AM NiF$_2$
where the orbital moment is approximately ten times larger than the spin moment \cite{hwlee2025}.
However, as shown in Table \ref{table2}, the total spin and orbital moments in \noo~
are nearly identical, in contrast to NiF$_2$.
In the OsO$_6$ octahedron, there are three different Os-O bond lengths \cite{nooKU},
but differences between them are only thousandths of Angstr\"oms,
resulting in $L_{\parallel}\approx L_\perp$.
So, the contribution of the anisotropic $g$-tensor in \noo~ is minimal,
since \noo~ is close to the ideal cubic perovskite and insulating.
As a result, the WM in \noo~ primarily arises from spin canting induced by the DMI,
as will be shown below.

Before closing this subsection, we mention that
atomic decomposition of the magnetic moments is approximate,
with different methods giving somewhat different values and perhaps implications. 
As mentioned above, we have calculated the Os1 (say, with spin-up) moment 
by two methods (Na gives up its electron, 
and O ions lie between antiparallel Os moments
and will have negligible induced moment). The moments on the other Os
sites are equal in magnitude but with different directions, according to the space
group. For Os1, {\sc vasp}, integrating the spin density within a sphere of
radius 1.413 \AA, gives the moment $(-0.058,-0.018,1.139)$$\mu_B$. From the
Os-centered Wannier functions, which employ no sphere but include 
interatomic effects, the moment is $(-0.052,-0.018,0.988)$$\mu_B$. 
The actual spin densities would differ by very little;
the differences arise only due to the (subjective) assignment to atoms.

\subsection{Magnonic properties}

\begin{figure}[tbp] 
 \includegraphics[width=\columnwidth]{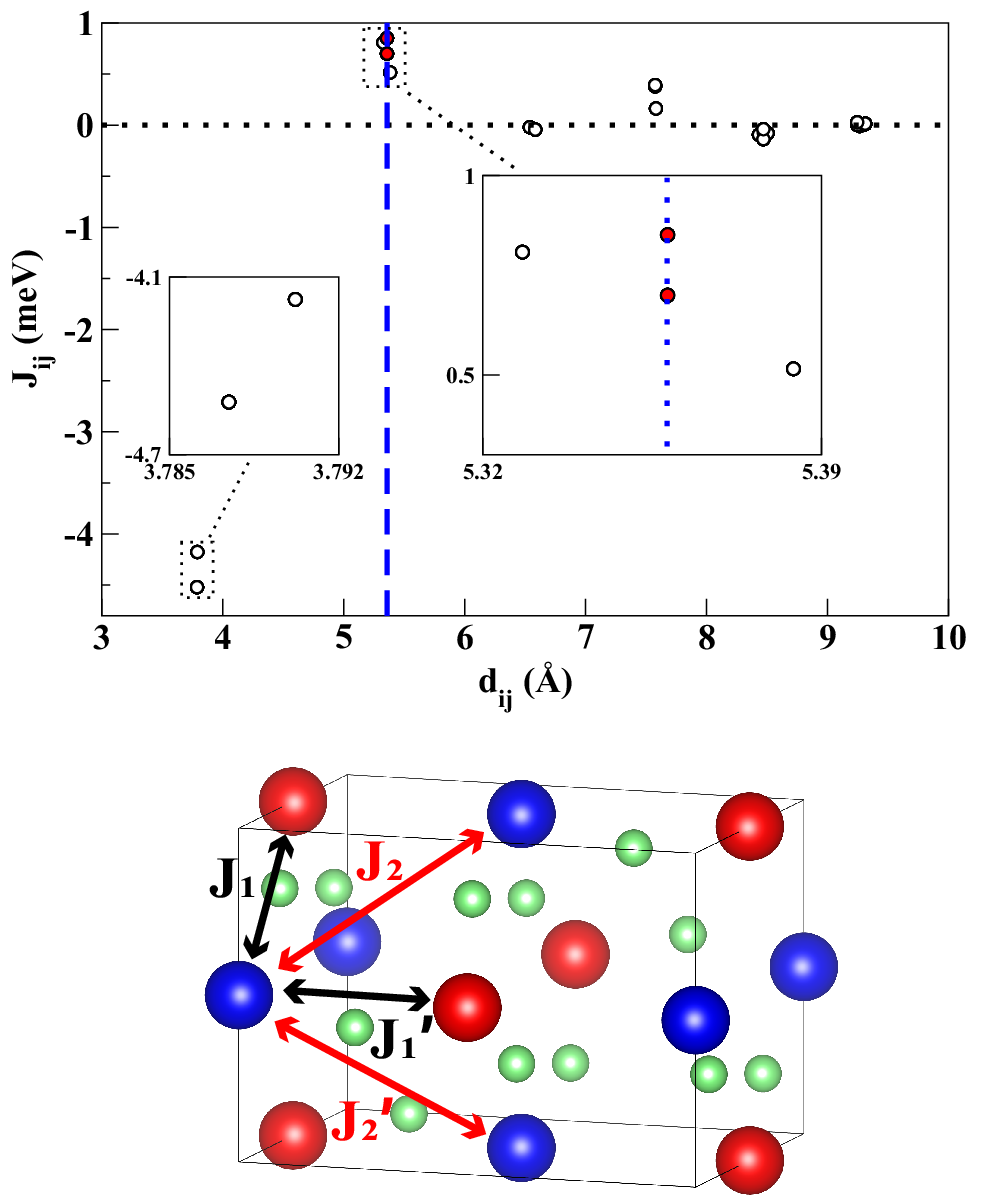}
\caption{Top: magnetic exchange parameters $J_{ij}$ versus the Os-Os distance $d_{ij}$,
obtained for the GGA+SOC(${\bf N}||\hat{z}$)+$U$ at $U_{eff}=0.5$ eV.
The Insets show (left) the two NN interactions separated by a tiny amount and (right) the four inequivalent NNN values. 
Note the difference in two ($J_2$ and $J'_2$) of the NNN values
(filled circles by the red color) at $d_{ij}\approx5.358$ \AA.
Bottom: origin of the inequivalent $J_2$ and $J'_2$ depicted in the G-type AFM cell.
In the cell, the large and small spheres indicate Os (red: spin-up, blue: spin-down) and O ions, respectively.
The Na ions are not shown, for easier visualization.
}
\label{exchange}
\end{figure}

The magnetic exchange parameters $J_{ij}$ for the 
the classical spin Hamiltonian $\mathcal{H}$ = $-\sum_{i,j}J_{ij}\mathbf{e}_i\cdot\mathbf{e}_j$
are calculated as implemented in the {\sc tb2j} code \cite{tb2j}.
Here, $\mathbf{e}_i$ = $\mathbf{S}_i/|S_i|$ is the normalized spin vector.
For simplicity, $J_{ij}$ is denoted below by $J_{j}$ for a fixed $i^{\rm th}$ Os ion.

Our calculated parameters with respect to the Os-Os distance are given in Fig. \ref{exchange}.
Due to the orthorhombic symmetry ($\frac{1}{\sqrt2}a\neq\frac{1}{2}b\neq \frac{1}{\sqrt2}c$), 
the two nearest neighbor (NN) interactions along the (101) and (010) directions differ only slightly,
as shown in the top panel of Fig. \ref{exchange}.
These inter-sublattice interactions at $d_{ij}\approx 3.8$ \AA~ 
lead to negative values of  $J_1(\approx J'_1)=-4.6 (-4.1)$ meV,
indicating that AFM order in \noo~ is induced by the NN  exchange interactions.
Around $d_{ij}\approx5.35$ \AA, four next nearest neighbor (NNN) interactions between intra-sublattices 
along the (100), (001), (111), and (11$\bar{1}$) directions
appear in the range of 0.5 meV -- 0.9 meV,
indicating spin-alignment NNN coupling that augments the NN antiferromagnetic coupling.
Higher order interactions are much smaller and affect only details of the magnon dispersion.

An unusual feature appears in two of the NNN interactions (denoted by $J_2$ and $J'_2$) 
along  the (111) and (11$\bar{1}$) directions
at $d_{ij}=\frac{1}{2}\sqrt{a^2+b^2+c^2}\approx5.358$ \AA,
as shown in the top panel of Fig. \ref{exchange}.
Along the paths of the two intra-sublattice interactions $J_2$ and $J'_2$ 
the environments of oxygen ions are  inequivalent
due to the crystallographic anisotropy of the AM, leading to a difference 
in the $J_2$ and $J'_2$ values by about 0.2 meV.
This is depicted in the bottom panel of Fig. \ref{exchange}.
This difference is small, about a fifth of that of  RuO$_2$ in magnitude \cite{chiralmagnon}, 
possibly because \noo~ is insulating. 
However, the difference is sufficient to lead to a unique feature of AMs arising
in the magnon spectrum.

\begin{figure}[tbp] 
 \includegraphics[width=\columnwidth]{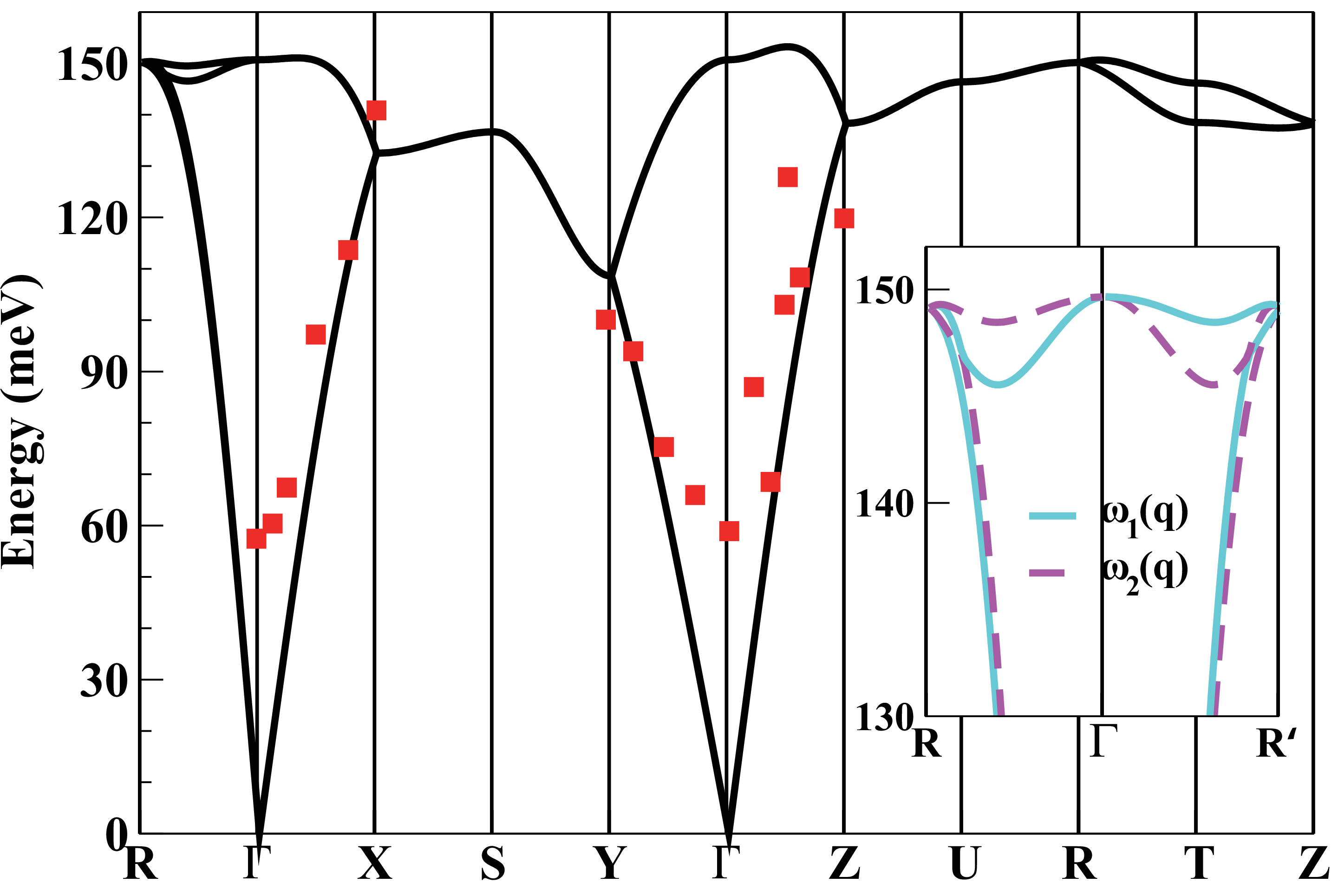} 
\caption{Magnon spectrum for the GGA+SOC(${\bf N}||\hat{z}$)+U at $U_{eff}=0.5$ eV. 
Here, the exchange integrals $J_{ij}$ are included up to the Os-Os distance $d_{ij}=6$ \AA. 
The AM character reflects in the bifurcation in the acoustic and optic modes along the $R-\Gamma$ line. 
The (red) squares denote the experimental values adapted from Ref. \cite{calder2017}. 
Inset: Blowup spectrum in the energy range above 130 meV along the $R(111)-\Gamma-R'(\bar{1}\bar{1}\bar{1})$ line 
where the bifurcations due to opposite chirality of the eigenvectors are clearly visible. 
}
\label{magnon}
\end{figure}

To calculate the magnon spectrum using the {\sc uppasd} code \cite{uppasd},
the exchange integrals are considered up to the Os-Os distance $d_{ij}=6$ 
\AA~ to include the $J_2$ and $J'_2$ values.
Figure \ref{magnon} shows the magnon spectrum, calculated without including
 a magnetic anisotropy term that shifts the $\Gamma$-point value to the 
experimental spin gap of $\sim$59 meV.
There are four branches, two acoustic and two optic, since this unit cell 
contains two pairs of magnetic sublattices.
Except for the spin gap, our calculated magnon curves 
are close to the measured values \cite{calder2017} 
that are depicted by symbols of the red squares in Fig. \ref{magnon}. 
Note especially that our maximum energy is very near the measured maximum (at the $X$ point).

What is specific to \noo~ is that the difference in the $J_2$ and $J'_2$ 
values is reflected on the bifurcation of the otherwise degenerate modes 
in both the acoustic and (more evident) optical branches along the $R-\Gamma$ line. 
As shown in the Inset of Fig. \ref{magnon} the spectrum along the $R-\Gamma-R'$ line
is split due to opposite chirality of the eigenvectors,
due to a symmetry constraint similar to the one discussed in Sec. IV \ref{spin-splitting}.
This has been
anticipated as a promising feature for the application of magnonics \cite{chiralmagnon}. 
Interestingly, the splittings are clearly distinguishable in both the 
acoustic and optical modes in the THz region,
whereas altermagnetic RuO$_2$ and MnTe have the splittings 
only in the acoustic mode \cite{chiralmagnon,liu2024}.
The largest splitting in \noo~ is about 5 meV in the optic mode around 150 meV (about 35 THz).
Note that the bifurcation along the $R-T-Z$ line is a trivial splitting 
between the optic modes.

\subsection{Dzyaloshinskii-Moriya interaction}

The two-ion Hamiltonian for the DMI term is given by
\begin{eqnarray}
{\cal H}_{DMI}= {\bf D}_{ij}\cdot[{\bf e}_i\times{\bf e}_j],
\label{eqn1}
\end{eqnarray}
where $D_{ij}=\|{\bf D}_{ij}\|$ is the parameter representing strength of the DMI.
(As above for $J_{ij}$, $D_{ij}$ will be denoted by $D_j$ for a fixed $i$-spin, as in $J_j$.)
We calculated ${\bf D}_{ij}$ from the total energy variation where 
the spin rotation is treated in linear response, implemented in the {\sc tb2j} package \cite{tb2j}.
In \noo, our calculations show that the NN ${\bf D}_{ij}$ constants
are dominant, with negligible longer range values.
(For longer range values, see the SM \cite{sm}.)
The NN values are, for the two bent bonds from octahedron rotation and tilting, 
${\bf D}_1$= (0.3753, --0.0722, 0) along the (101) direction
and ${\bf D}_2$= (0.1903, 0, --0.2091) along the (010) direction, in units of meV.
So, $D_1$=0.38 meV and $D_2$=0.28 meV, about 10\% of the corresponding exchange parameters.
Using a formula due to Jackeli and Khaliullin \cite{jackeli} for the canting angle 
$\phi=\frac{1}{2} {\rm tan}^{-1} (\frac{D_i}{J_i})$, $\phi\approx2.3^\circ$,
close to the value obtained from our calculated spin moment vectors in Sec. IV \ref{WM}.

Our calculated values of ${\bf D}_{ij}$ are smaller (by several tenths of meV) than those reported by Ntallis {\it et al.} \cite{ntallis2021} 
The differences may be due to choices of computational methods. For the DFT part, 
Ntallis {\it et al.} applied a fully relativistic, full potential linear muffin-tin method implemented 
in the {\sc rspt} code \cite{rspt}. Thereafter, different codes were used to obtain the
various constants and the spin wave spectrum.
Although differences among various approaches are common, 
there are two additional possible reasons for the relatively small discrepancy in this case.

Spin-orbit interactions were treated differently: 
at the first level, we used a SOC approach originally implemented 
in the linear augmented plane method \cite{wep1980}, 
whereas Ntallis {\it et al.} adopted a fully relativistic scheme. 
While these two methods typically yield quite similar results for 5$d$ elemental solids \cite{wep1980}, 
subtle differences can still arise. 
Second, the DFT+U schemes employed in the two studies may have been different. 
Different choices of this functional can result in somewhat different 
results \cite{kwlee2006,ylvisaker2009}, however the nearly (locally) cubic spin density of
Os would show less difference than if the moment had low symmetry spin density.
The larger values reported by Ntallis {\it et al.} led to an overestimation of 
the top magnon frequencies \cite{ntallis2021}, 
which occur at the zone boundary. 
Our results show good agreement for the maximum of the spin wave spectrum, 
note specifically the $X$ and $Y$ points. 
Whatever the source of the differences, it is satisfying
that the ratios of $\frac{D_1}{J_1}$ obtained both computational methods are similar.

\begin{figure}[tp] 
\includegraphics[width=\columnwidth]{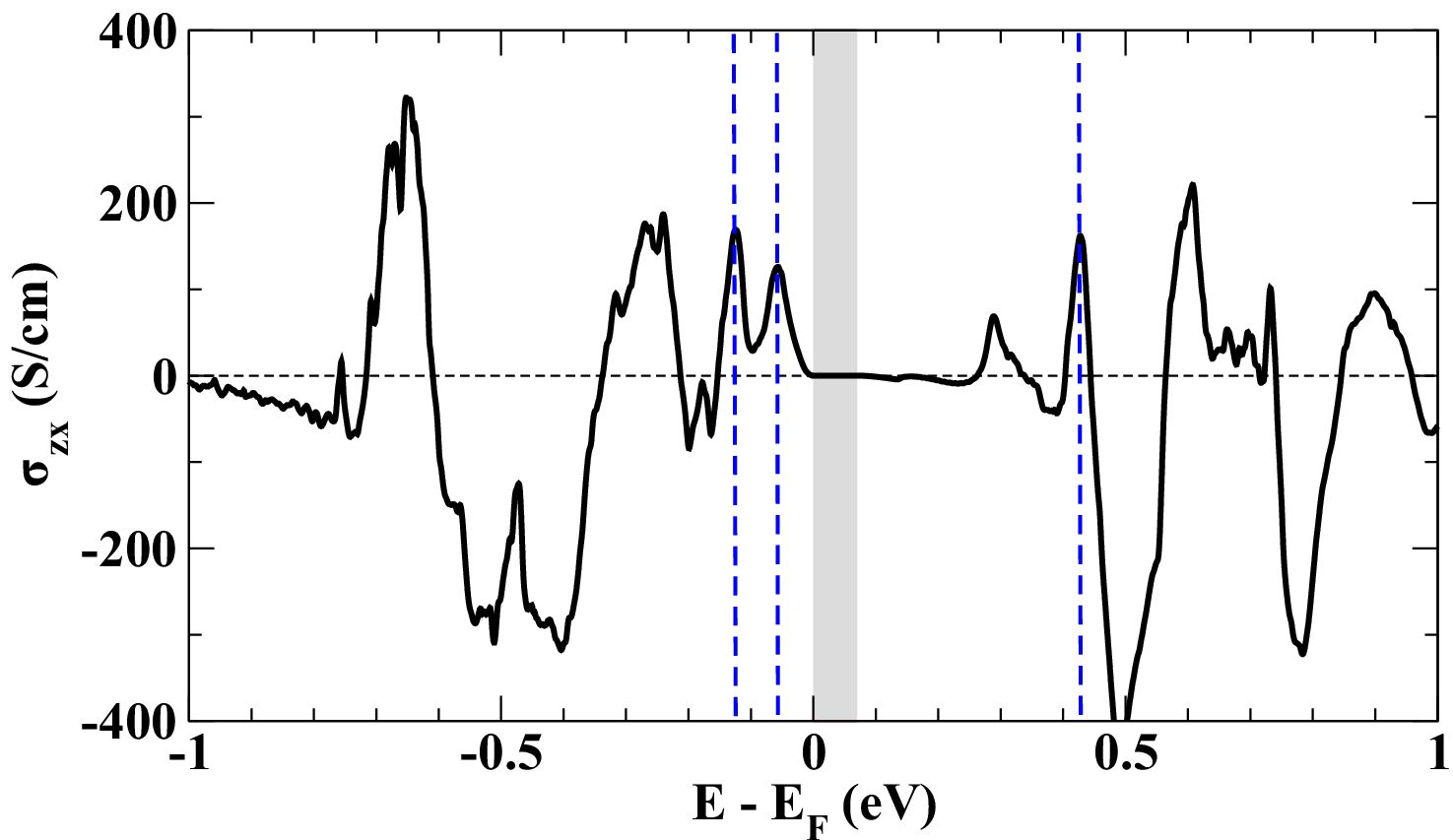} 
\caption{Anomalous Hall conductivity $\sigma_{zx}$ for GGA+SOC(${\bf N}||\hat{z}$)+$U$ at $U_{eff}=0.5$ eV. 
For the case of ${\bf N}||\hat{z}$, $\sigma_{xy}=\sigma_{yz}=0$ by the magnetic space group symmetry
(for details, see text).
The (blue) dashed lines indicate doping levels of 0.29 holes, 0.11 holes,
and 0.62 electrons (per f.u.), from left to right in order, estimated from the rigid band approximation.
The zero value in the energy gap region shaded here indicates a topologically trivial insulating state.
}
\label{ahc}
\end{figure}

\subsection{Anomalous Hall effect}
\label{AHE}
In an AM material, separation of majority and minority characters of bands 
in some regions of BZ can lead to the anomalous Hall effect \cite{ahe_rev2022}, 
as observed in ferromagnetic metals.
As discussed in Sec. IV \ref{WM}, by the symmetry of the magnetic space group 
the Hall vector ${\bf h}$ connecting the applied electric field to
the Hall current density is allowed only along the $\hat{z}$ and $\hat{y}$ directions 
for the case of ${\bf N}\|\hat{y}$ and $\|\hat{z}$, respectively, 
whereas ${\bf h}$ is zero for ${\bf N}\|\hat{x}$.
Thus for our case of ${\bf N}\|\hat{z}$ nonzero anomalous Hall effect exists only for $\sigma_{zx}$.

The AHC, calculated using GGA+SOC+U for the 
easy axis ${\bf N}\|\hat{z}$, is given in Fig. \ref{ahc} versus band filling
in the range --1 eV to 1 eV.
Several peaks with a large magnitude appear,  
 with a maximum value of 450 S/cm at 0.5 eV above $E_F$, 
implying electron doping on the Na site could achieve this value.
Below $E_F$, peaks with magnitude of 150 -- 200 S/cm at $-0.06$ eV and $-0.12$ eV 
could be reached by doping of 0.11 holes and 0.29 holes per f.u., 
estimated by the rigid band approximation, possibly achieved by Na vacancies.
These features remain nearly unchanged at $U_{eff}=1$ eV,
while the peaks below $E_F$ are somewhat enhanced at $U_{eff}=2$ eV,
as given in SM \cite{sm}.
This approximation would provide a reasonable estimate, 
since there is no Na character near $E_F$ (see SM \cite{sm}).
These values of AHC are much larger than those calculated for MnTe \cite{ahe2023}.
Note that the {\it ab initio} AHC values obtained from the Berry curvature 
provide a reasonable tendency, although such estimates can show quantitative 
disagreement with experimental values \cite{feng2022}.

\section{Discussion and Summary}
Our work supports that \noo~is an example of an unusual small gap insulating altermagnet,
whose G-type AFM order and spin dynamics are dominated by first and
second neighbor exchange couplings.
The quite weak net magnetization arises from Dzyaloshinskii-Moriya canting 
due to low symmetry bonding, aided by the altermagnetic ordering. 
To answer one question posed in the Introduction, our DFT work supports that
\noo~ lies in a crossover regime between Slater and Mott insulator.
As mentioned in the Introduction, this regime of a gap opening/closing
MIT is more involved than the conventional Slater and Mott pictures, 
although we will continue to consider here this separation.

The distortion into the gadolinium orthoferrite structure already
creates an unexpected separation of valence and conduction bands.
The combination of SOC with a small Hubbard $U$ repulsion steadies 
the (T=0) gap, seemingly supporting the Mott picture.
These features, relying on the small gap (or not), may instead reflect a small deficiency of 
the GGA functional. Studies of excitations or properties upon doping, which
differ between Slater and Mott systems, could clarify this crossover regime. 

Our values of the spin exchange parameters led to excellent agreement
with the experimental magnon dispersion in the high energy region. 
The resulting spinwave spectrum provides the novel example of a chiral magnon
in the optic branch.  
With the anisotropy term obtained elsewhere, the magnon spectrum including
the chiral anomaly is obtained quantitatively. 
Upon doping, \noo~ should show a substantial anomalous Hall effect.

Although emphasizing that our results give support to the picture that the magnetic transition in NaOsO$_3$ is intermediate between Slater- and Mott-types (neither clearly applying) and accepting the usual characterization as a continuous volume transition (which favors the Slater picture), our work has not focused unduly on experimental details of the transition. However, in the data plotted \cite{johannes2023} by Zhou {\it et al.} in their Fig. 2 (using the data adapted from Ref. \cite{calder2012}), a very small, but above stated experimental uncertainty, volume change occurs at the transition. Our estimate is $\Delta V/V$$\approx$2$\times 10^{-4}$. This discontinuity indicates an (extremely) weakly first order transition, which would likely entail a discontinuous opening of the gap (our calculations address only T=0 electronic and magnetic structure). This feature might support the characterization of Zhou {\it et al.} as a ``muted Mott transition''. Such a small discontinuity, if true,  may rival the smallest volume collapse transition ever reported, and might affect the analysis in terms of critical exponents and indeed in concept of such a delicate distinction between characters of magnetic ordering and gap opening.

A weakness of the above picture, in line with muted Mott transition picture,
is that the seeming (very small) discontinuity in volume is not clear in the individual structural
parameters, and indeed the Slater picture has been the conclusion of earlier
works \cite{shi2009,calder2012,savrasov,moha2018,gurung2018}, in spite of the need to
apply a modest onsite repulsion $U$ to obtain a gap.
The relevance of our work to this question has not been great, except to point out so far
neglected complexities where experimental
data become even more relevant. Shi {\it et al.} have shown \cite{shi2009} that both the resistivity and
magnetic susceptibility display an {\it abrupt} increase at $T_{MIT}$, not necessarily
the signature of a continuous transition involving smooth onset of magnetic order
and gap opening. Given the large size of the (low temperature) ordered 
Os moment $\sim 1\mu_B$, and the increasing susceptibility   
as $T$ is lowered in the conducting phase above $T_{MIT}$ (linear behavior of 
$1/\chi(T)$ indicative of spin fluctuations \cite{shi2009}), 
and the abruptly increasing susceptibility above the transition, the transition will have spin
disorder-order character, an aspect not included in either the Slater or Mott pictures.

The measured resistivity also raises related questions. A continuous opening of the gap
and of magnetic order could suggest less abrupt behavior, the gap opening decreasing
conductivity sharply, while magnetic order reduces 
spin fluctuation scattering proportional to the order parameter
while leaving phonon scattering unchanged. Here the altermagnetic
character of \noo~ may become relevant, because both hole and electron bands bordering
the gap are spin polarized and will enable spin scattering separate from that
of the magnons. In addition, the electron band broadening due to phonons and
spin fluctuations will be noticeable and have its own implications, 
especially for gap opening. 

The case of Cd$_2$Os$_2$O$_7$, another Os$^{5+}$ osmate \cite{y.feng2021}, 
presents an example of another unusual MIT, {\it smooth} in the 
resistivity and Hall response. The smoothness of the transition would not
allow any discontinuous volume change. The underlying
magnetic behavior in this pyrochlore structure osmate is quite different from
that of G-type AFM \noo.
It seems likely that there is not yet any viable theory for
this intermediate region between Slater and Mott ordering in a material specific manner.

\section{Acknowledgments}
We acknowledge useful communications 
with Young-Joon Song for comments on perovskite altermagnets,
with Hyun-Woo Lee for discussion on anisotropy of orbital moments in altermagnets, 
with Changyoung Kim for spin splitting in the altermagnets,
with Bongjae Kim for various DFT calculations of \noo, and 
with M. D. Johannes on the character of the magnetic transition. 
H.S.C and K.W.L were supported by National Research Foundation of Korea (NRF) Grant (RS-2024-00392493),
and M.C.J. was supported by NRF of Korea Grant No. NRF2019R1A2C1009588.
K.H.A. was supported by Grant No. 25-17490S of the Czech Science Foundation
and No. CZ.02.01.01/00/22\_008/0004594 (TERAFIT) of the Czech Ministry of Education, Youth and Sports.

\vskip 8mm
{\bf End of Paper}

\newpage\onecolumngrid
\renewcommand\thefigure{S\arabic{figure}} 
\newcounter{myfig} 
\setcounter{figure}{0}

\renewcommand\thetable{S\arabic{table}} 
\newcounter{mytable} 
\setcounter{table}{0}

\section{Supplemental Material}


This Supplemental Material provides additional information to provide support to the
text. The information is described in the figure captions.
\vskip 8mm

\begin{figure*}[!ht] 
\includegraphics[scale=0.15]{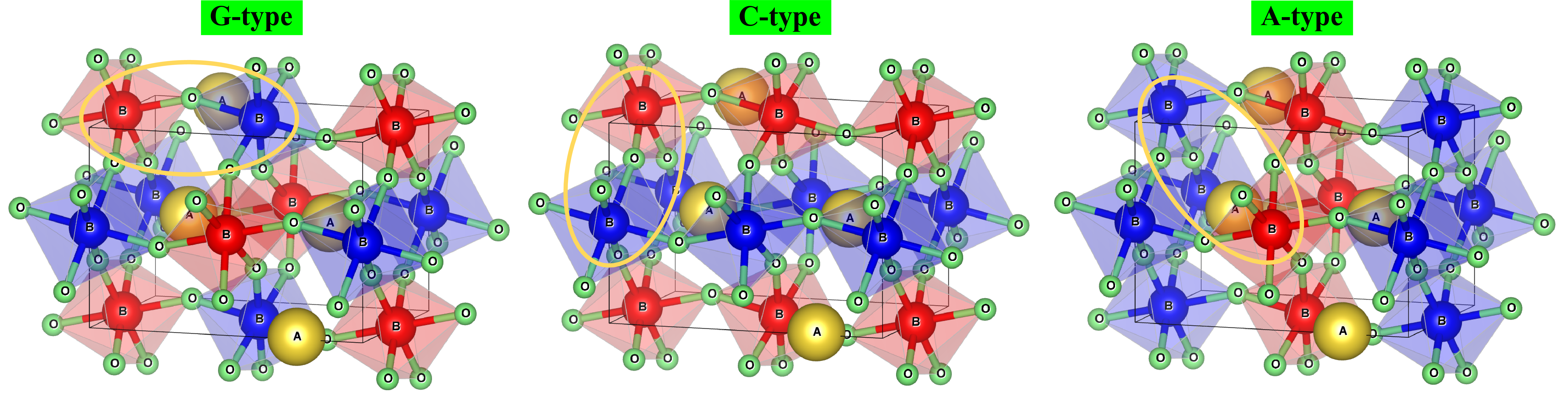}
\caption{Three types of collinear antiferromanetic order 
in the GdFeO$_3$-type (gadolinium orthoferrite) perovskite structure.
Each magnetic sublattice, marked by brown ellipses, is connected
by a combination of two-fold rotational $C_{2j}$
and non-primitive translation \tr~ magnetic symmetries: 
$[C_2||C_{2y}t_{(0\frac{1}{2}0)}]$, 
$[C_2||C_{2z}t_{(\frac{1}{2}0\frac{1}{2})}]$, and
$[C_2||C_{2x}t_{(\frac{1}{2}\frac{1}{2}\frac{1}{2})}]$ 
for the  G-, C-, and A-types, respectively.
A fundamental result is that these perovskites 
are altermagnets in each of these orderings.
}
\label{afmstr}
\end{figure*}

\begin{figure*}[!ht] 
\includegraphics[scale=0.20]{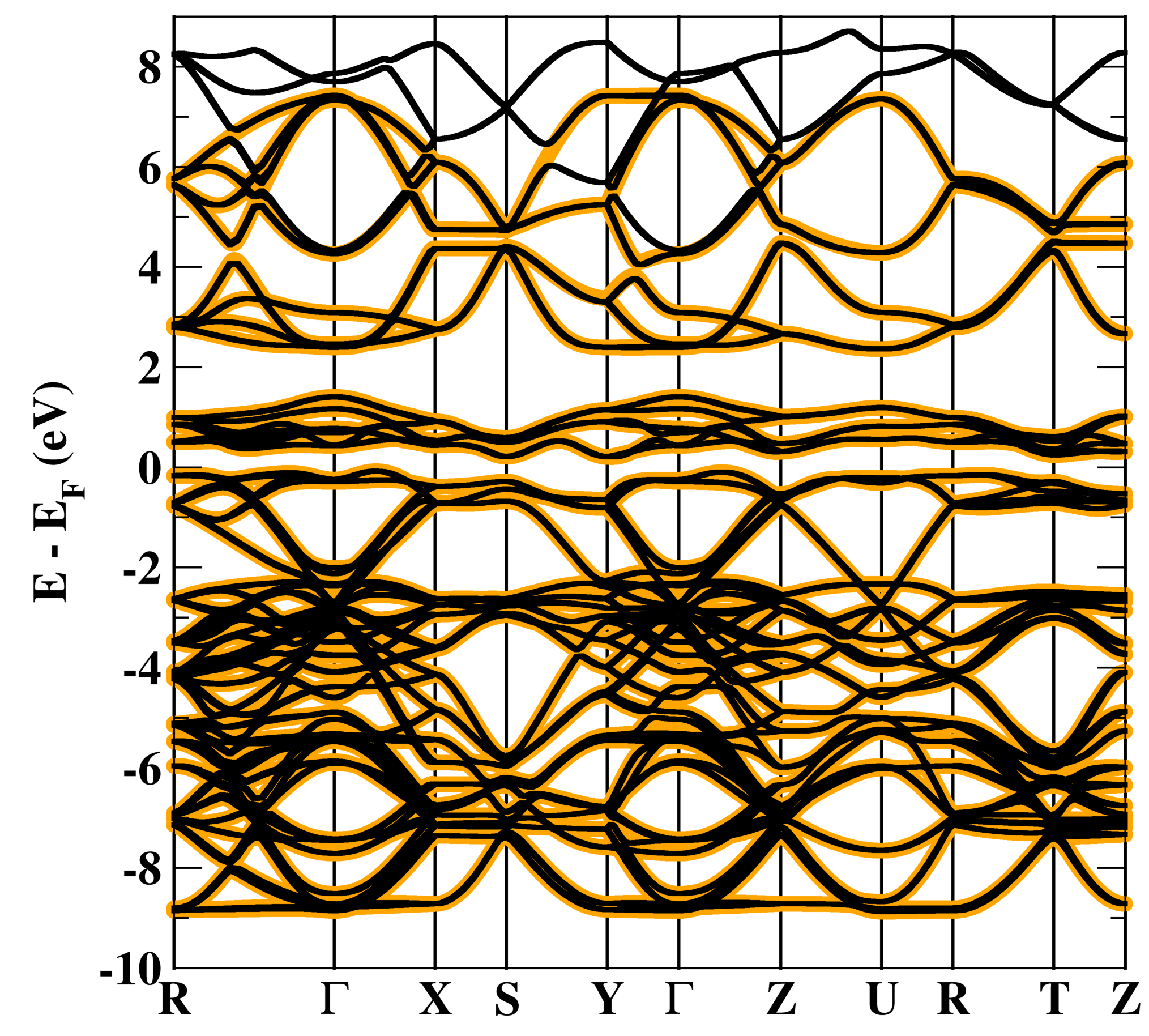}
\caption{Comparison of the Wannierized bands (brown symbols) with the 
	{\it ab initio} bands, from the GGA+SOC(${\bf N}\|\hat{z}$)+U at $U_{eff}=0.5$ eV,
	showing the very good representation over a 15 eV span. 
	All Os-$d$ and O-$p$ orbitals were used in the Wannier basis set.
}
\label{wan}
\end{figure*}

\begin{figure*}[!ht] 
\includegraphics[scale=0.30]{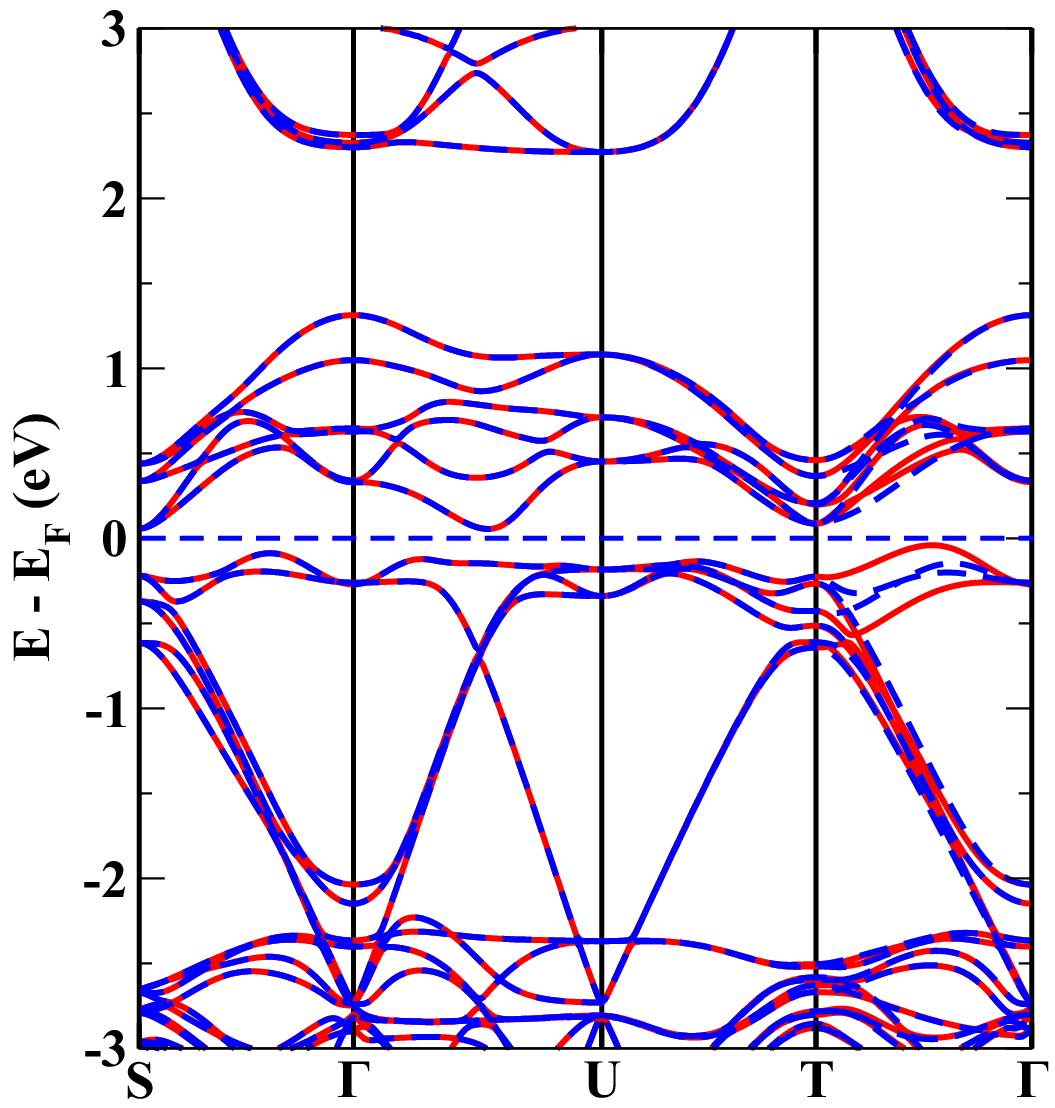}
\caption{GGA band Structure along the $S-\Gamma-U-T-\Gamma$ line 
showing spin-splitting inside the Brillouin zone.
Here, the spin characters are colored by red (spin-up) and blue (spin-down).
}
\label{bs_inside}
\end{figure*}

\begin{figure*}[!ht] 
\includegraphics[scale=0.23]{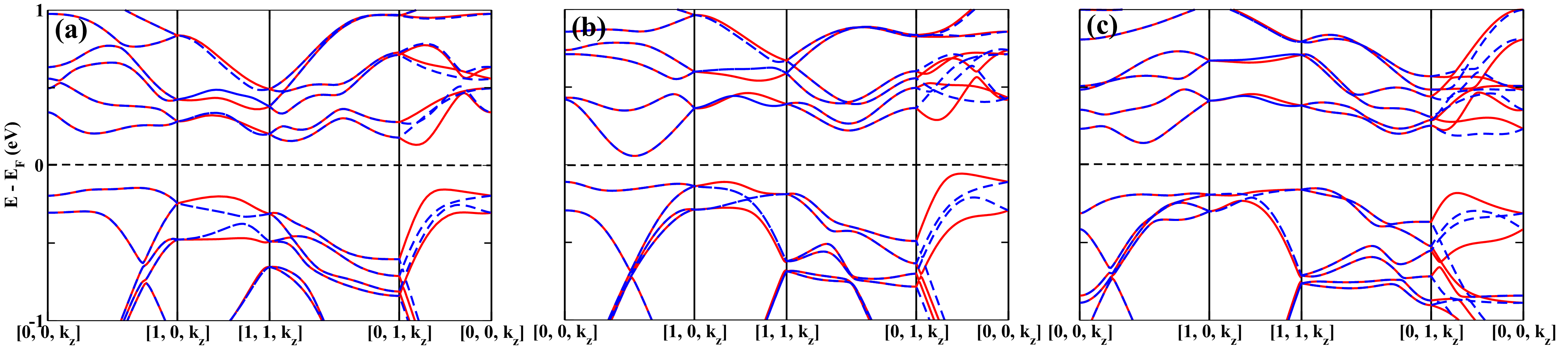}
\caption{Fine scale band structures, within the GGA level, plotted along the 
	line surrounding the zone-boundary over the range of --1 eV to 1 eV. 
	These are on $k_x-k_y$ planes with values $k_z=\frac{1}{4}$, $\frac{1}{2}$, 
	and $\frac{3}{4}$ (units of $\frac{\pi}{c}$), in frames (a), (b), (c) respectively.
	The red-solid (blue-dashed) lines indicate spin-up (down) bands. 
	Spin-splittings occur along the lines parallel to the $k_x=0$ line
	on the each plane with a fixed $k_z$. 
	The size of the splittings is clearly visible in the right hand panels of each frame.
}
\label{bs_kz}
\end{figure*}

\begin{figure*}[!ht] 
	\includegraphics[scale=0.30]{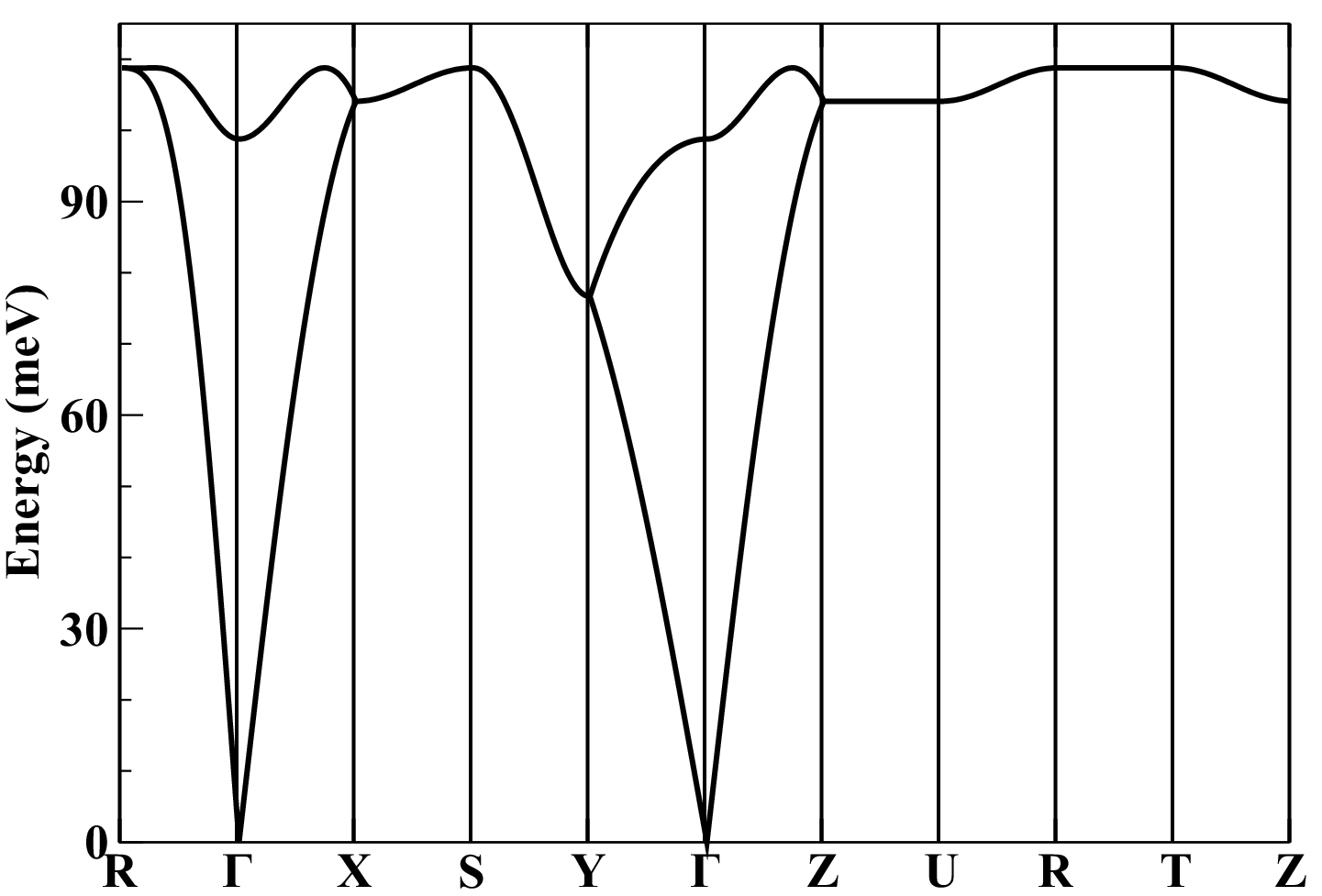}
\caption{Magnon spectrum from the GGA+SOC(${\bf N}||\hat{z}$)+U at $U_{eff}=0.5$ eV,
	including only the NN exchange integrals.
	This demonstrates that chiral splitting arises from 
	the inequivalent NNN $J_2$ and $J'_2$, which are not considered for this plot.
	Note that the trivial splitting between two optic branches 
	along the $R-T-Z$ line is about 0.3 meV, invisible here.
}
\label{magnon4}
\end{figure*}

\begin{figure*}[!ht] 
	\includegraphics[scale=0.30]{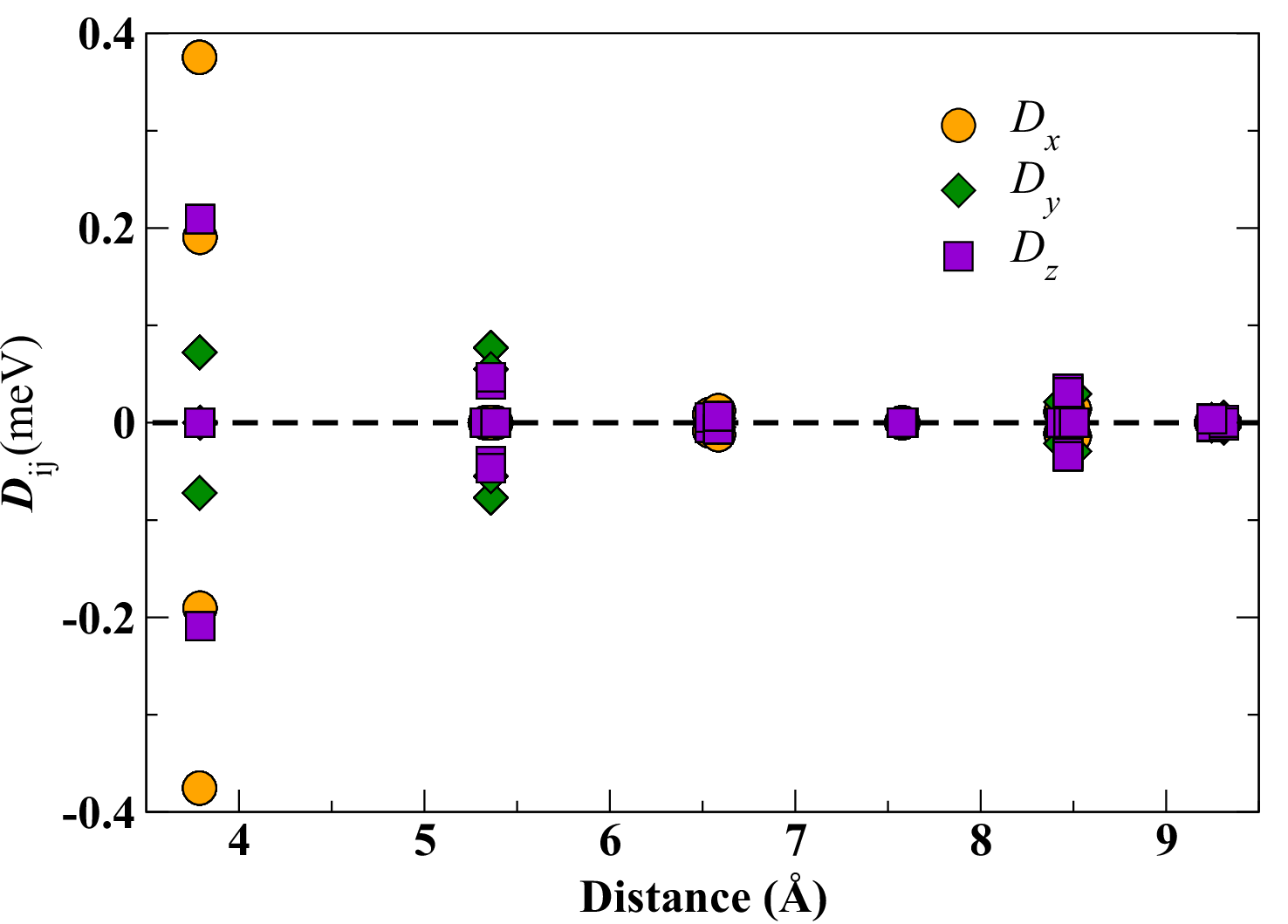}
\caption{Each component of the Dzyaloshinskii-Moriya interaction(DMI) vectors ${\bf D}_{ij}$ (meV) 
	versus the Os-Os distance $d$ (\AA),
	indicating that the NN DMIs are dominant. 
	These are obtained from the GGA+SOC(${\bf N}\|\hat{z}$)+U 
	calculations at $U_{eff}=0.5$ eV. 
}
\label{dmi}
\end{figure*}

\begin{figure*}[!ht] 
\includegraphics[scale=0.30]{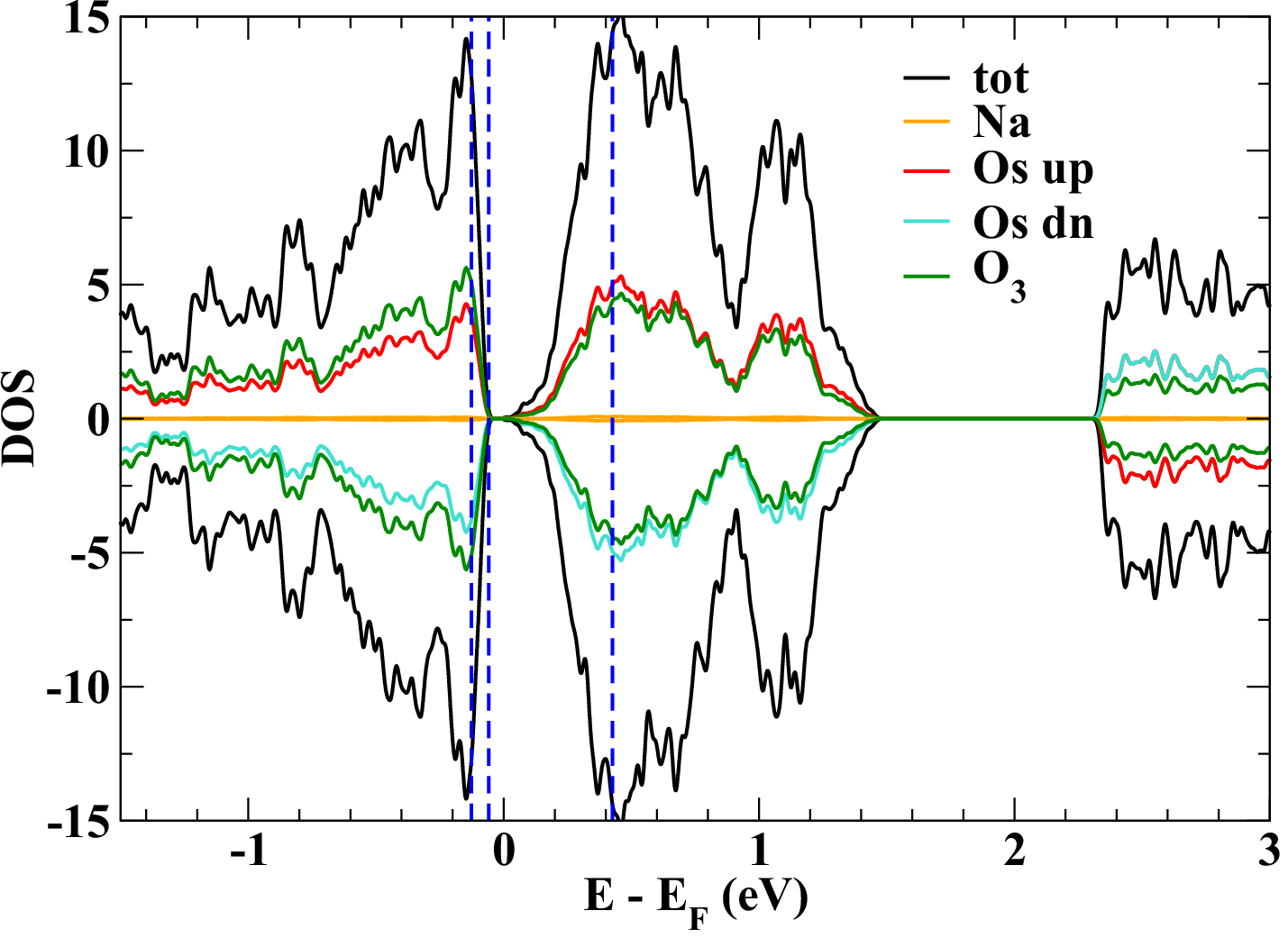}
\caption{Total and atom-projected densities of states from the GGA+SOC(${\bf N}\|\hat{z}$)+U at $U_{eff}=0.5$ eV, in units of states per eV per f.u.
	The vertical (blue) dashed lines denote the doping levels mentioned in the caption of Fig. \ref{ahc}.
}
\label{dos}
\end{figure*}

\begin{figure*}[!ht] 
	\includegraphics[scale=0.30]{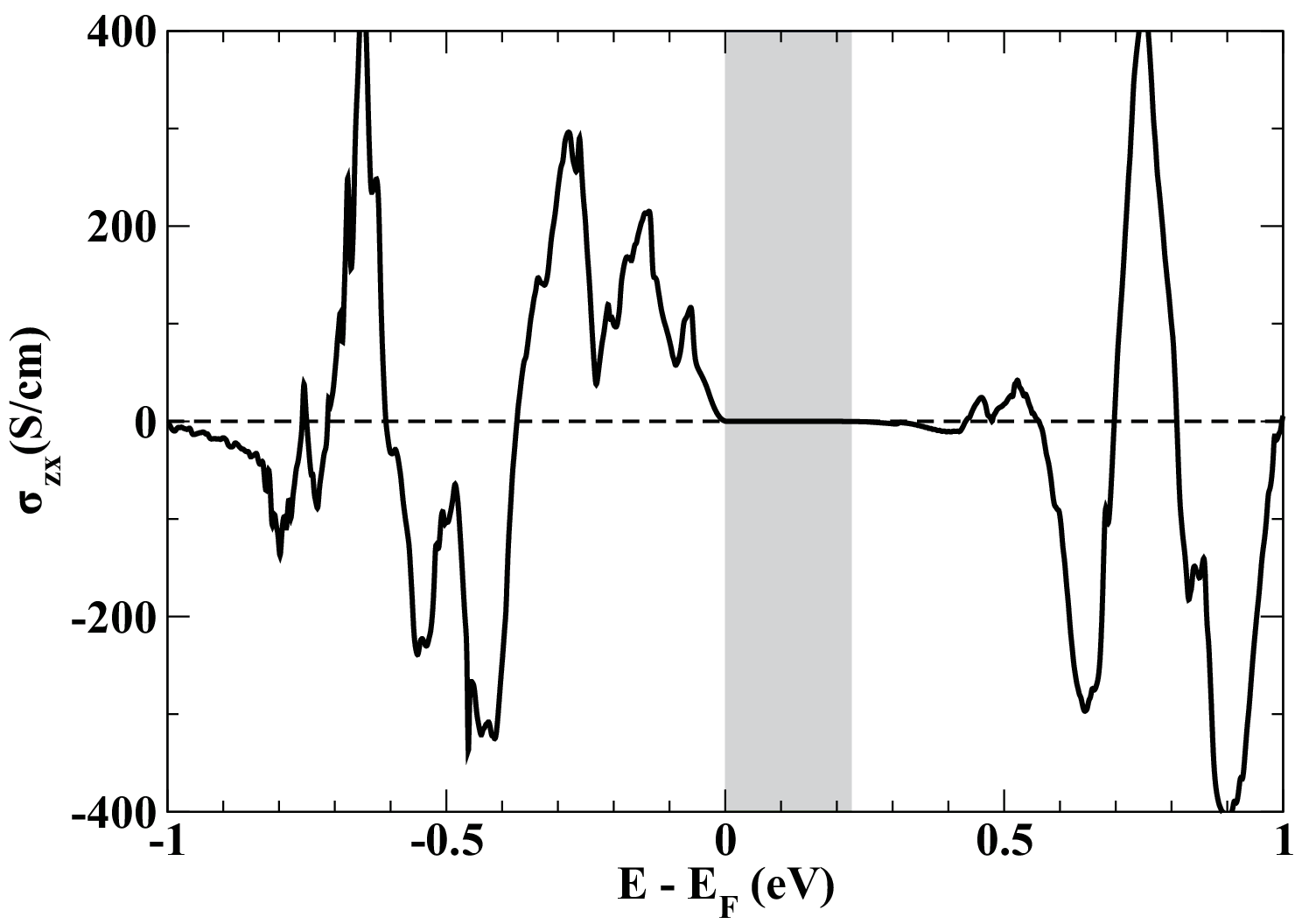}
    \hskip 8mm
    \includegraphics[scale=0.30]{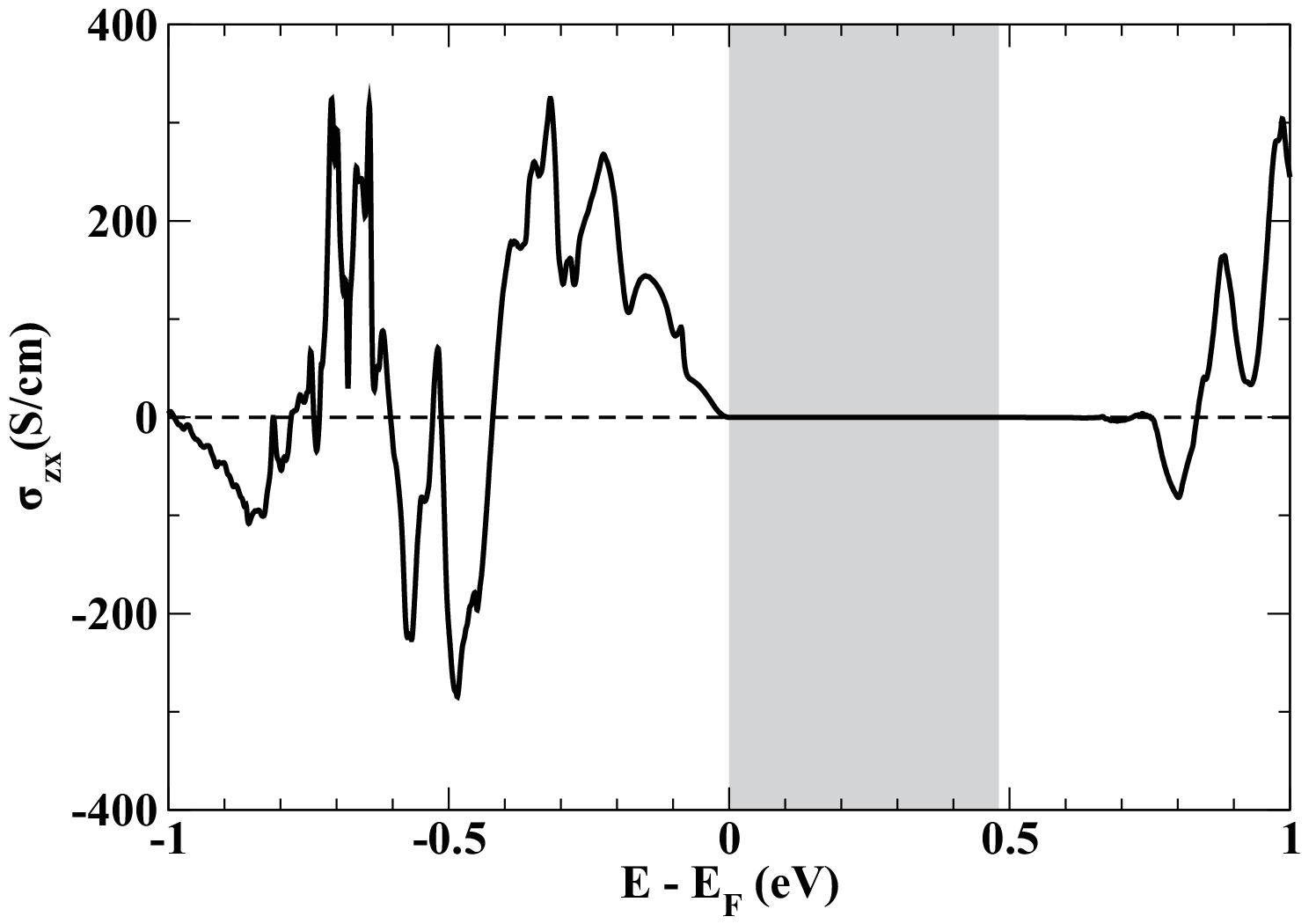}
\caption{Anomalous Hall conductivity $\sigma_{zx}$ for GGA+SOC(${\bf N}\|\hat{z}$)+$U$ at $U_{eff}=$1 (left) and 2 (right) eV.
In particular, the peaks located between --0.3 eV and $E_F$ are robust against variations of $U$ value studied here.
The shaded area denotes the zero value in the energy gap region.
}
\label{AHE_U}
\end{figure*}

\begin{figure*}[!ht] 
	\includegraphics[scale=0.30]{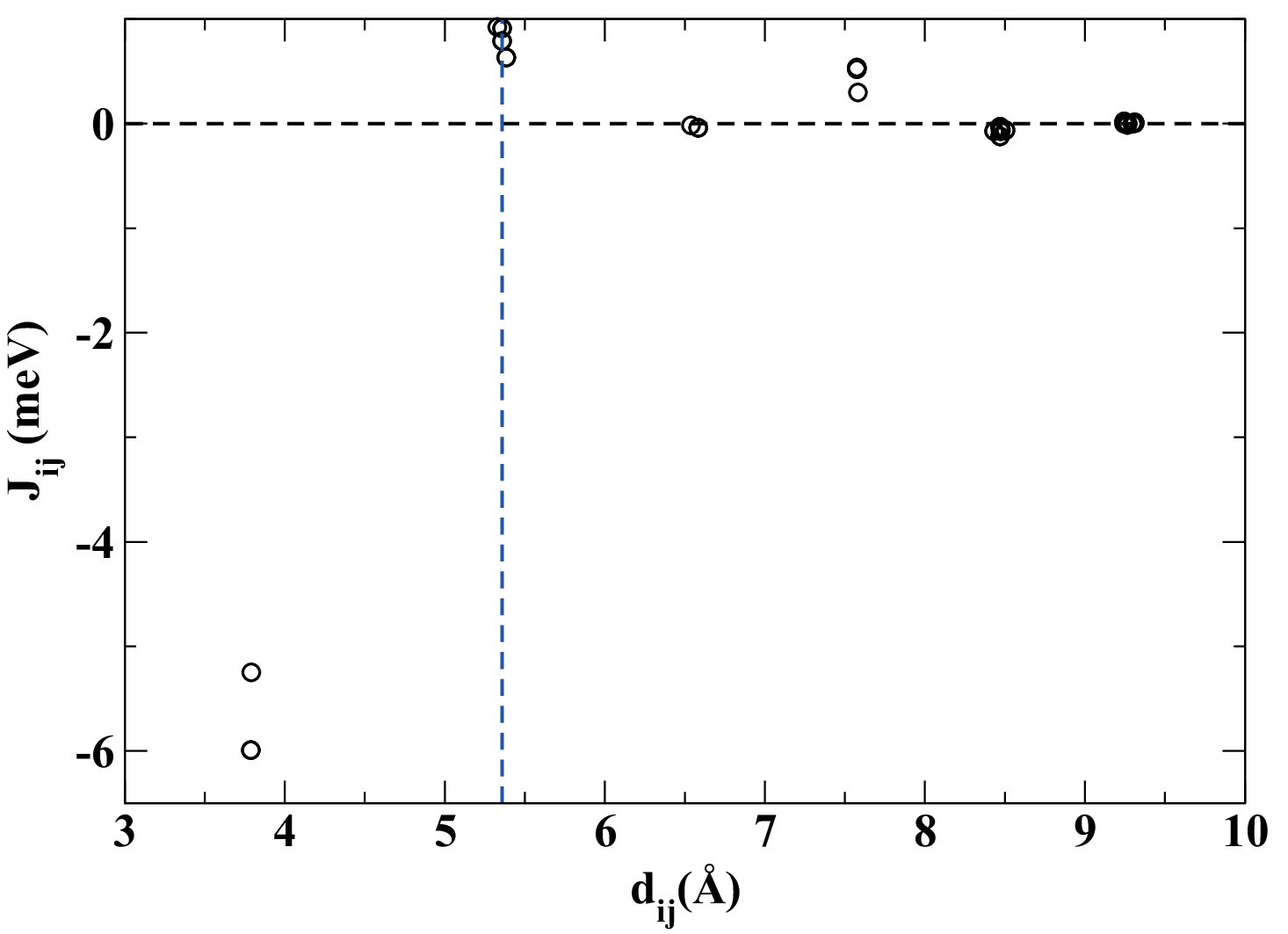}
    \hskip 8mm
    \includegraphics[scale=0.30]{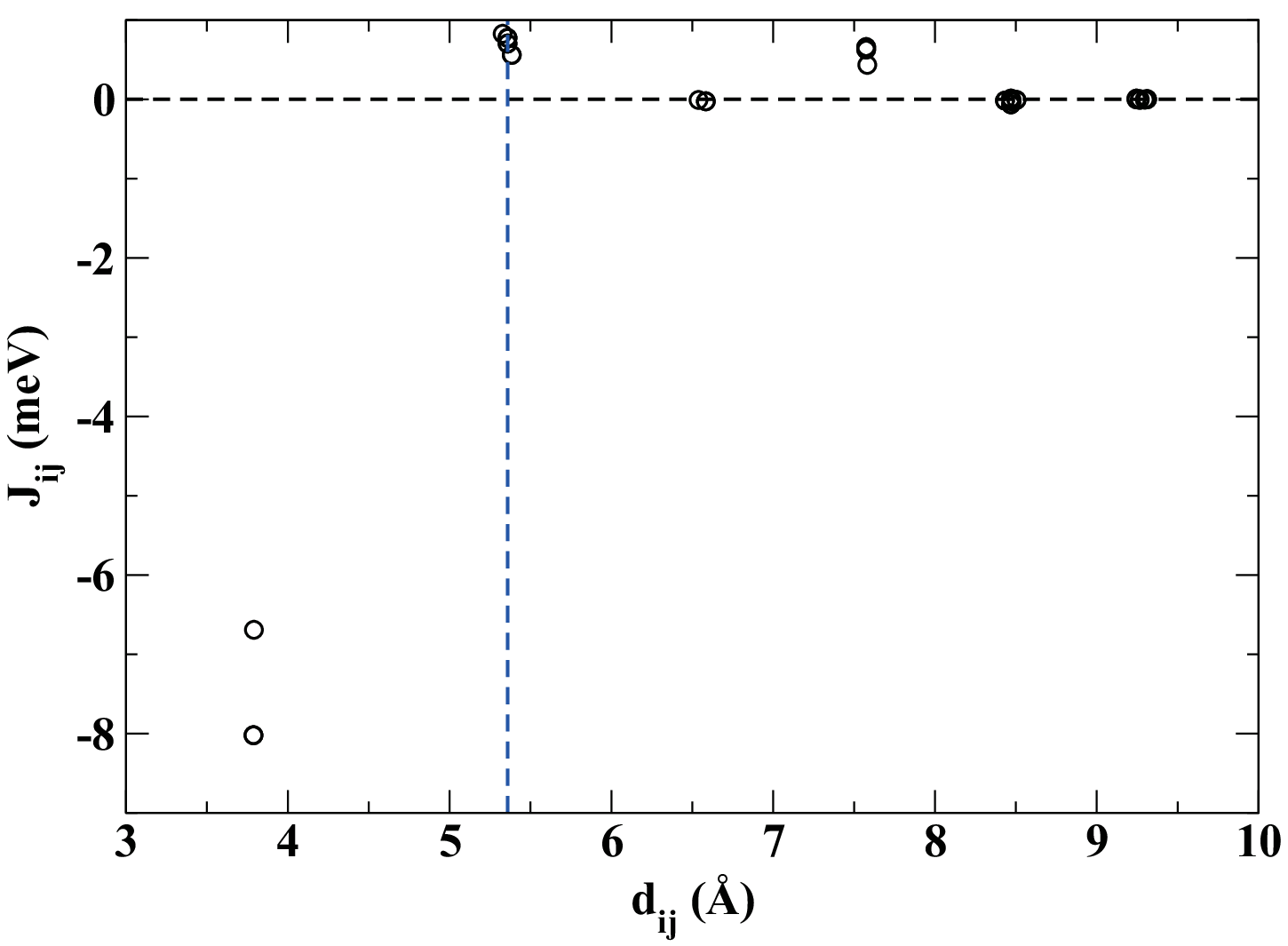}
\caption{
Magnetic exchange parameters $J_{ij}$ versus the Os-Os distance $d_{ij}$ for GGA+SOC(${\bf N}\|\hat{z}$)+$U$ at $U_{eff}=$1 (left) and 2 (right) eV.
The difference in the $J_2$ and $J'_2$ values, located at the vertical dashed-line, is by about 0.12 (0.07) meV at $U_{eff}=$1 (2) eV, invisible here.
Note that the magnitude of the NN exchange parameters increases with increasing $U_{eff}$ due to the enhancement of the Os local magnetic moments: 0.989, 1.130, and 1.338 (in units of $\mu_B$) for $U_{eff}=$0.5 eV, 1 eV, and 2 eV, respectively.
This enhancement leads to an increase in the maximum energy of the magnon spectrum, as shown below.
}
\label{exchange_U}
\end{figure*}

\begin{figure*}[!ht] 
	\includegraphics[scale=0.40]{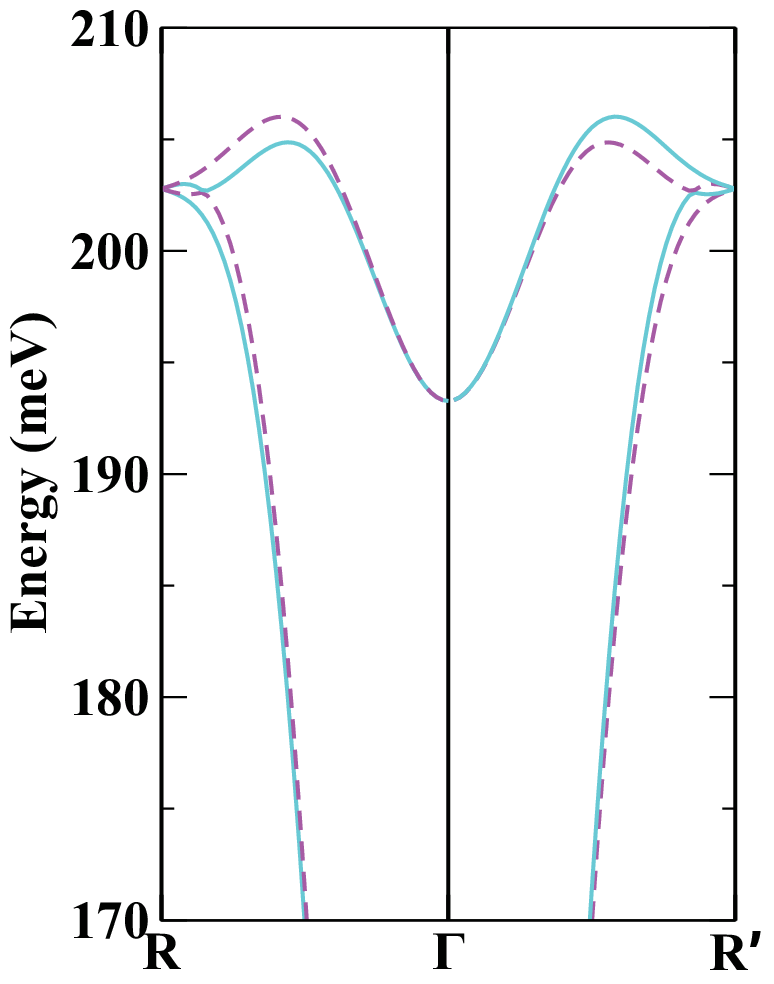}
    \hskip 8mm
    \includegraphics[scale=0.40]{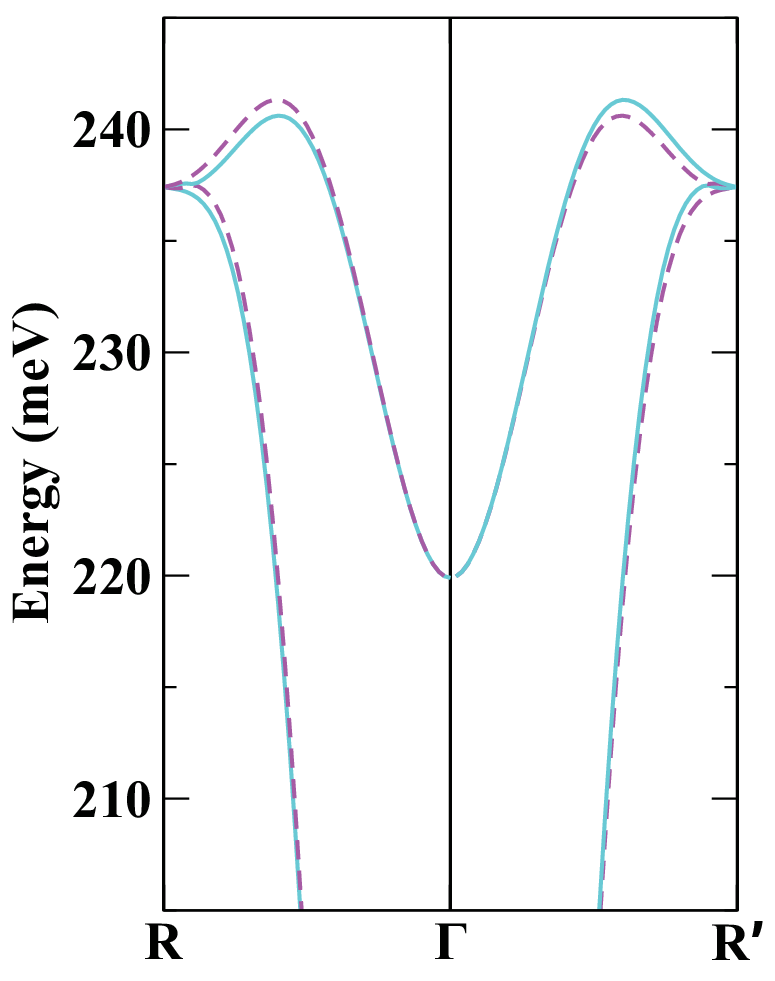}
\caption{
Enlarged magnon spectrum around the optic modes along the $R(111)-\Gamma-R'(\bar{1}\bar{1}\bar{1})$ line where the splittings due to opposite chirality of the eigenvectors are clearly visible, 
for GGA+SOC(${\bf N}\|\hat{z}$)+$U$ at $U_{eff}=$1 (left) and 2 (right) eV.
The largest splitting in the optic modes is about 2 (1) meV at $U_{eff}=$1 (2) eV.
However, the maximum energy for these values of $U_{eff}$ is significantly larger than that of the experimental value given in the main text.
}
\label{magnon_U}
\end{figure*}

\begin{table*}[!ht]
\begin{center}
\caption{Occupation matrices for the Os $5d$ orbitals of \noo~ in GGA.
	These are represented in terms of the real harmonics, implemented in the 
        all-electron full-potential {\sc fplo} program [K. Koepernik and H. Eschrig, Phys. Rev. B {\bf 59}, 1743 (1999)]. 
        The ccupation matrices provide the trace of 3.114 (1.811) for the majority (minority) orbitals,
	resulting in about 4.915 electrons, significantly different from the $5d^3$ configuration of 
	Os$^{5+}$. This difference of real charge from formal charge is a well documented feature
	of transition metal oxides; for Os, the real charge is near 5 electrons for even higher formal valences (7+, 6+). The unexpected differences between the occupation matrix
	values and what would normally be expected of the majority $t_{2g}^3$ occupations are 
	because the natural axes for the OsO$_6$ octahedron are rotated away from the global coordinate
	system in which the occupation matrices are expressed.
}
\vskip 2mm 
\begin{tabular}{cccccccccccc}\hline\hline
      atom &\multicolumn{5}{c}{Os1}&~&\multicolumn{5}{c}{Os2} \\\cline{2-6}\cline{8-12}  
ortibal& $xy$ & $yz$ & $z^2$ & $xz$ & $x^2-y^2$&~& $xy$ & $yz$ & $z^2$ & $xz$ & $x^2-y^2$\\\hline
     ~& 0.3935 & 0.1778 & 0.0357&--0.0420 & 0.1118 &~& 0.6024& 0.2530& 0.0304&--0.0882 &0.0890\\
     ~&  0.1778 & 0.7574 & 0.0027 & 0.0694 &--0.0672 &~& 0.2530 & 0.5485 &--0.0430& 0.0195&--0.0775\\
majority&  0.0357& 0.0027& 0.5329&--0.2769&--0.0768 &~& 0.0304&--0.0430& 0.8406&--0.0957&--0.0756\\
 ~& --0.0420& 0.0694&--0.2769& 0.6200&--0.0245 &~&  0.0882& 0.0195&--0.0957& 0.4162&--0.2034 \\
~&  0.1118&--0.0672&--0.0768&--0.0245& 0.8102 &~&  0.0890&--0.0775&--0.0756&--0.2034& 0.7063 \\\hline

~&0.2966 & 0.0436& 0.0211&--0.0236& 0.0543 &~& 0.3729& 0.0710& 0.0155&--0.0293& 0.0128\\
~& 0.0436& 0.4141& 0.0198& 0.0108&--0.0106 &~& 0.0710 &0.3379&--0.0175& 0.0148&--0.0528\\
minority &  0.0211& 0.0198& 0.3101&--0.0707&--0.0225 &~&  0.0155&--0.0175& 0.4315&--0.0134&--0.0119\\
~& 0.0236 &0.1080&--0.0707& 0.3737& 0.0029 &~& --0.0293& 0.0148&--0.0134& 0.3050&--0.0638\\
~& 0.0543&--0.0106&--0.0225& 0.0029& 0.4167 &~& 0.0128&--0.0528&--0.0119&--0.0638& 0.3639\\
\hline\hline
\end{tabular}
\end{center}
\end{table*}

\end{document}